\documentclass[usenatbib]{mn2e}

\usepackage{graphicx}

\newcommand{\arepo}{{\sc arepo}}
\newcommand{\gadget}{{\sc gadget}}
\newcommand{\kpc}{{\rm kpc}}
\newcommand{\h} {{\rm H}}
\newcommand{\hi} {{\rm H}\,{\small\rm I}}
\newcommand{\hii} {{\rm H}\,{\small\rm II}}
\newcommand{\kms} {{\rm km~s}^{-1}}
\newcommand{\cm} {{\rm cm}^{-3}}
\newcommand{\cmsq} {{\rm cm}^{-2}}
\newcommand{\pc} {{\rm pc}}
\newcommand{\Mpc} {{\rm Mpc}}
\newcommand{\mo}{{\rm M}_\odot}
\newcommand{\Gyr}{{\rm Gyr}}
\newcommand{\K}{{\rm K}}
\newcommand{\gsim}{\lower.7ex\hbox{$\;\stackrel{\textstyle>}{\sim}\;$}}
\newcommand{\lsim}{\lower.7ex\hbox{$\;\stackrel{\textstyle<}{\sim}\;$}}

\newcommand{\FM}[1]{#1}

\defcitealias{Marinacci2013}{MPS14} 
\defcitealias{Planck2013}{Planck collaboration} 

\title[Diffuse gas and metals in disc simulations]
{Diffuse gas properties and stellar metallicities in cosmological
simulations of disc galaxy formation} 
\author[F.~Marinacci et al.]
{Federico Marinacci$^{1,2}$\thanks{E-mail:
federico.marinacci@h-its.org}, R\"udiger~Pakmor$^1$, Volker~Springel$^{1,2}$ \newauthor 
  and Christine M. Simpson$^1$\vspace*{0.2cm}\\
  $^1$Heidelberger Institut f\"{u}r Theoretische Studien,
  Schloss-Wolfsbrunnenweg 35, 69118 Heidelberg, Germany\\
  $^2$Zentrum f\"ur Astronomie der Universit\"at Heidelberg,
  Astronomisches Recheninstitut, M\"{o}nchhofstr. 12-14, 69120
  Heidelberg, Germany}

\date{Accepted 2014 June 6.  Received 2014 June 5; in original form 2014 March 19}
  
\begin{document}

\pagerange{\pageref{firstpage}--\pageref{lastpage}}
\pubyear{2014}

\maketitle

\label{firstpage}

\begin{abstract}
  We analyse the properties of the circum-galactic medium 
and \FM{the metal content} of the stars comprising
the central galaxy in eight hydrodynamical `zoom-in' simulations of
disc galaxy formation. We use these properties as a benchmark for
our model of galaxy formation physics implemented in the moving-mesh
code \arepo, which succeeds in forming quite realistic late-type
spirals in the set of `Aquarius' initial conditions of Milky Way-sized
haloes. Galactic winds significantly influence the morphology of the
circum-galactic medium and induce bipolar features in the distribution
of heavy elements. They also affect the thermodynamic properties of
the circum-galactic gas by supplying an energy input that sustains its
radiative losses. Although a significant fraction of the heavy
elements are transferred from the central galaxy to the halo, and even
beyond the virial radius, enough metals are retained by stars to
yield a peak in their metallicity distributions at about $Z_{\odot}$. 
All our default runs overestimate the stellar [O/Fe] ratio, an effect
that we demonstrate can be rectified by an increase of the
adopted SN type Ia rate. Nevertheless, the models have difficulty in
producing stellar metallicity gradients of the same strength as
observed in the Milky Way.
\end{abstract}


\section{Introduction}\label{sec:intro}

On large scales, dark matter only cosmological simulations \citep[such
  as the Millennium runs,][]{Millennium, MillenniumII} have been very
successful in reproducing the observed clustering properties of matter
in the Universe, and have been of decisive help in establishing the
current $\Lambda$CDM cosmological paradigm.  Numerical simulations
have also become an important tool for studying how galaxies in the
high-redshift and present-day Universe form and evolve.  However, at
galactic and sub-galactic scales, the general picture of galaxy
formation is still fraught with uncertainty.  Much of the current
simulation work in the field therefore attempts to clarify the role of
different physical processes operating on galactic and sub-galactic
scales, as well as their interplay within the cosmological context.

A major difficulty in galaxy formation simulations is the
intrinsic multi-scale and multi-physics nature of the problem. In
fact, baryonic physics must be properly taken into account in addition
to gravity if one wants to obtain a realistic description of galaxy
formation and evolution. Unfortunately, important physical processes
acting on baryons (e.g.~AGN accretion) operate on spatial and temporal
scales that are very much smaller than the galaxy as a whole.
Nevertheless, these processes play a fundamental role in shaping global
galactic features. It is this disparity of scales, together with the
uncertainties in our understanding of the baryonic physics involved, that
makes galaxy formation such a challenging problem. 

To overcome these difficulties, a heuristic parametrization of the
most important physical mechanisms for the galaxy formation process is
frequently adopted (see \citealt{Vogelsberger2013}, and references therein).
Recently, significant efforts have also been made to
develop models that try to self-consistently include more of the
relevant physical processes and hence to push back the use of ad hoc
prescriptions to ever smaller scales \citep{Hopkins2011, Hopkins2013, Renaud2013}. 
Treating much of the baryon physics of star
formation and associated feedback as `subgrid' is however still the
most widely used approach in the field and is inevitable in large
cosmological boxes for computational reasons. It has allowed
significant advances in our understanding of galaxy formation by
identifying the most relevant physical processes and isolating their
role in determining the observed galaxy properties.

In particular, substantial progress has been made on
the formation of late-type disc galaxies similar to the Milky Way
\citep{Agertz2011, Brooks2011, Guedes2011, Aumer2013b, Stinson2013,
Marinacci2013}. For decades, this has been a sore point in the
context of the $\Lambda$CDM cosmology because the galaxies predicted
by simulations were at odds with what was observationally
known. The primary cause of these discrepancies is the so-called
\textit{overcooling} problem \citep{Balogh2001}: baryons at the
centres of the dark matter haloes that host galaxies cool efficiently
through radiative processes on a very short time scale, invariably
leading to the formation of overly massive and overly concentrated
galaxies \citep[e.g.][]{Abadi1, Scannapieco2008, Scannapieco2009, Stinson2010}.

Therefore, realistic late-type galaxies can only be formed in
simulations when the gas is supplied with an adequate amount of energy
capable of partly offsetting its fast cooling rate and curtailing the
production of stars. The source of this energy ultimately lies in the
back-reaction that the star formation process (and also AGN accretion)
exerts on the gas. Several channels (e.g.~radiative, thermal, kinetic)
are in principle conceivable to couple this energy to the cooling gas,
with their relative importance still being debated
\citep[e.g.][]{Sales2013}. Different authors agree, however, that the
adoption of very strong feedback physics lies at the heart of recent
successful simulations of the formation of spiral galaxies
\citep{Guedes2011, Aumer2013b, Stinson2013, Marinacci2013}.

Independent of the detailed mechanism responsible for the efficient
coupling of feedback energy to the star-forming gas, strong stellar
feedback is expected to drive galactic-scale winds that expel
metal-enriched gas out of the star-forming disc, possibly to distances
comparable to or beyond the virial radius of the enclosing dark matter
halo. This theoretical expectation appears to be confirmed by
observations of star-forming galaxies at high and low redshift,
showing that outflows from the sites of star formation to the
circum-galactic medium (CGM) are a common phenomenon
\citep{Pettini2000, Strickland2000, Martin2002, Strickland2004, Soto2012}.

Simulations suggest that the ejection of metal-enriched gas from
galaxies to their haloes via galactic winds has at least two important
consequences for galactic evolution. First, the ejected gas is (at
least temporarily) unavailable for star formation, thereby preventing
its rapid conversion into stars and alleviating the problem of forming
overly massive bulge-dominated galaxies. Much of the ejected gas
remains however bound to the galaxy's dark matter halo and thus can be
re-accreted, providing fuel for late-time star formation 
that progressively builds up the disc \citep[][]{Scannapieco2012}. 

Second, simulated wind material is in general more
metal-rich with respect to the diffuse circum-galactic gas, and
therefore represents an important source of metals for this gas
component. Besides affecting its chemical composition, the interaction
of the wind with the CGM can also significantly alter the dynamical
and thermodynamical state of the diffuse gas. As we mentioned above,
the mixture of wind material and circum-galactic gas provides fuel for
late-time star formation. Therefore, galactic outflows also have an
important influence on the properties of stars (in particular their
metallicities), which in turn are responsible for wind generation.
Therefore, the properties of the CGM and the metal content of stars
comprising the central galaxy offer an important and powerful
opportunity to check the consistency of the galaxy formation physics
models adopted in numerical simulations.

In this paper we thus aim at investigating how the galactic-scale
outflows contained in the simulations of \citet[][hereafter
MPS14]{Marinacci2013} affect the features of the CGM surrounding
late-type galaxies and those of the stars that they contain. We
use this information as a test of our galaxy formation physics model --
in which the galactic wind strength has been primarily set to reproduce the
expected stellar mass  but \textit{not} the physical state and metal
enrichment of the CGM or the detailed metal content of the stellar
component -- to better understand the limits of our approach, with the
goal of using the simulated results to improve its underlying
physics description. To this end, we examine the properties of these
two components, focusing in particular on their metal content, in a
suite of eight hydrodynamical simulations of disc galaxy formation in
Milky Way-sized haloes carried out with the moving-mesh code \arepo\
\citep{Arepo}. The simulation set has been previously introduced in
\citetalias{Marinacci2013}, where we analysed the stellar mass
distribution of the simulated galaxies, showing that our numerical
methodology, which has been designed for applications
in large-scale cosmological simulations \citep{Illustris}, 
produces realistic disc galaxies that agree well with the observed scaling 
relations of late-type systems.

The paper is structured as follows. In Section~\ref{sec:simulations},
we summarize the salient characteristics of our simulation set and
those of the numerical techniques that we used to perform the runs. We
present the results of our analysis in Section~\ref{sec:results}
separately for gas (Sec.~\ref{sec:gas}) and stellar
(Sec.~\ref{sec:stars}) properties. Finally, we discuss our findings in
Section~\ref{sec:discussion} and draw our conclusions in
Section~\ref{sec:conclusions}.

\section{The simulation set} \label{sec:simulations}

We analyse a sub-set of the cosmological simulations of disc galaxy
formation discussed in \citetalias{Marinacci2013}. These simulations can be
viewed as hydrodynamical counterparts of the dark matter only
simulations carried out in the Aquarius project \citep{Springel2008},
and include eight haloes (labeled Aq-A, Aq-B, etc., until Aq-H)
selected to have a final total mass in a small range around $\sim
10^{12}\,\mo$, comparable to recent determinations of the mass of the
Galaxy \citep{Sakamoto2003, Battaglia2005, Dehnen2006, Xue2008,
  Li2008, Boylan-Kolchin2013}.  In addition to the mass selection
criterion, the Aquarius haloes were chosen to have a relatively
quiet merger history by enforcing a moderate isolation criterion,
which favours the formation of extended, star-forming discs. More
details about the selection criteria and the properties of these
haloes can be found in the original Aquarius paper
\citep{Springel2008} and in \citet{Boylan-Kolchin2010}.

The Aquarius simulation suite uses the so-called ``zoom-in'' technique
to follow the formation of a target halo with much higher resolution
than achievable in a homogeneously sampled cosmological box. In
practice, the Lagrangian region that will collapse to form the
central structure is discretized with a high number of low mass
particles surrounded by several layers of progressively more massive
particles that fill the total simulation volume. The masses of these 
low-resolution particles increase with distance from the central object,
such that the regions further away from the high-resolution volume are
sampled more coarsely but still sufficiently well to ensure an
accurate accounting of all tidal forces acting on the target
halo. This saves computational time without sacrificing the accuracy
of the calculation.

In this paper we only analyse a single resolution level which is
denoted as level '5', following the naming scheme of the Aquarius
project. This also corresponds to the default resolution of the
simulations presented in \citetalias{Marinacci2013}. The main
properties of this resolution level\footnote{The quoted mass
  resolution only applies exactly to the Aq-C-5 halo.  There are slight
  differences for the other haloes of the simulation set, within a
  factor of 2. The detailed list of properties for all the haloes can
  be found in Table 1 of \citetalias{Marinacci2013}.} are a
characteristic baryonic mass resolution of $\simeq 4.1\times
10^5\,\mo$, a dark matter mass resolution of $\simeq 2.2\times
10^6\,\mo$ and a maximum gravitational softening length at $z \leq 1$
of $680\,\pc$ in physical units for all mass components (dark matter
and baryons) in the high-resolution region. In comoving units, the
softening length is kept fixed for $z > 1$ in the high-resolution
region and its physical counterpart is thus growing until $z = 1$,
where it reaches the maximum allowed value and is then held
fixed. Common to all the resolution levels in the Aquarius suite are a
periodic box of $100\,h^{-1}\Mpc$ on a side and a $\Lambda$CDM
cosmology with parameters $\Omega_{\rm m} = 0.25$, $\Omega_{\rm b} =
0.04$, $\Omega_{\rm \Lambda} = 0.75$, $\sigma_{8} = 0.9$, $n_{\rm s} =
1$ and a Hubble parameter $H_{0} = 73~\kms\Mpc^{-1}$ (hence implying
$h = 0.73$). This set of cosmological parameters is the same that
  has been used in the Millennium and Millenium-II simulations
  \citep{Millennium, MillenniumII} and is consistent with WMAP-1
  results.  Compared to the more recent Planck findings
  (\citetalias{Planck2013} \citeyear{Planck2013}), haloes of given
  mass in a WMAP-1 cosmology tend to form slightly earlier and hence
  have on average slightly higher concentration.  In
  $10^{12}\,{\rm M}_{\odot}$ haloes, the increase in the average concentration
  is $\sim 10-20$\% \citep[see][]{Maccio2008}.

To include gas in the original Aquarius initial conditions, we split
each original dark matter particle, regardless of whether it is
located in the high- or low-resolution region, into a pair composed of
a dark matter particle and a gas cell, so that the whole simulation
volume is filled with gas. We note that this differs from many studies
performed with the SPH technique \citep[e.g.,][]{Stinson2010,
  Guedes2011} where gas is often added to the high-resolution region
only, causing a pressure discontinuity at the transition between the
high- and low-resolution regions. The mass ratio in each dark matter
particle-gas cell pair is set by the cosmic baryon mass fraction,
while their positions and velocities are assigned by requiring that
the centre-of-mass and the centre-of-mass velocity of the pair are
preserved.  This is done by displacing the dark matter and cell
distributions such that two interleaved particle distributions are
formed that are locally shifted relative to each other by half of the
original mean inter-particle spacing.

We evolve the initial conditions with the moving-mesh cosmological
code \arepo\ \citep{Arepo}. This code solves the gravitational and
collision-less dynamics by using the same TreePM approach employed in
the popular SPH code \gadget\ \citep{Springel2005b}.  For what
concerns the hydrodynamic part, \arepo\ solves Euler's equations on an
unstructured Voronoi mesh by adopting a finite-volume
discretization. The flux exchanges between cells are computed via a
second-order MUSCL-Hancock scheme \citep[e.g.,][]{Toro1999} coupled to
an exact Riemann solver. The special trait of \arepo\ compared to
Eulerian grid-based codes is that the set of points generating its
unstructured Voronoi mesh are allowed to freely move, and this
additional degree of freedom is usually exploited by moving the
mesh-generating points with the local fluid velocity. This causes a
transformation of the mesh that adapts itself to the characteristics
of the flow, giving rise to a quasi-Lagrangian numerical method which
is manifestly Galilean-invariant and keeps the mass of each gas cell
approximately constant. Further details about the code structure can
be found in \citet{Arepo}, while a detailed comparison between the
moving-mesh approach and the smoothed particle hydrodynamics technique
is presented in the tests performed by \citet{Vogelsberger2012} and
\citet{Sijacki2012}. Extensions of the basic method to include ideal
magnetohydrodynamics have also been developed \citep{Pakmor2013} and
applied successfully in cosmological simulations \citep{Pakmor2014}.

The present set of simulations uses a comprehensive model for galaxy
formation physics \citep{Vogelsberger2013} that implements the most
important baryonic processes. The model has been specifically developed for the
\arepo\ code and was calibrated to reproduce a small set of key
observational findings for the global galaxy population, such as
the cosmic star formation history and the galaxy stellar mass
function. The full description of the implementation is given in
\citet{Vogelsberger2013} to which we also refer for a discussion of
the predicted galaxy properties at $z = 0$ \citep[see also][for an
analysis of these properties at high redshift]{Torrey2013}. Among the
primary features of the model are:

\begin{figure*}
\resizebox{17.8cm}{!}{\includegraphics{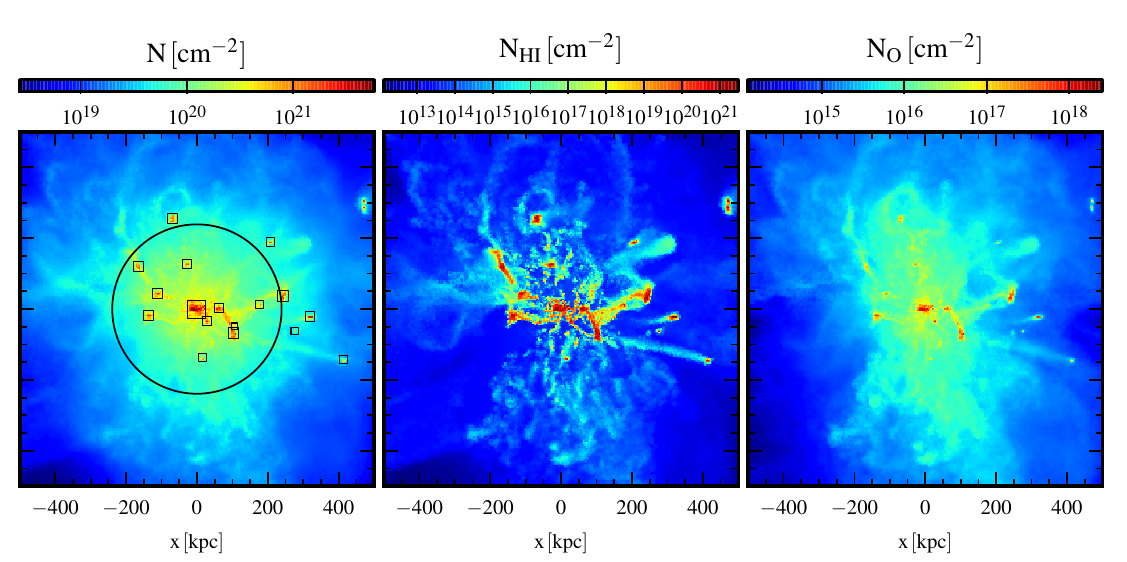}}
\caption{Column density maps of different chemical species for the
Aq-A-5 simulation at $z = 0$. To compute the column density only the
diffuse gas (i.e., non star-forming gas cells) has been taken into
account. In the figure, the halo is oriented such that the central
galaxy is seen edge-on to assess the impact of the adopted
galactic wind model on the morphology and the metal enrichment of the
CGM. To create the maps no cut in systemic velocity
has been applied, but only the gas with a distance along the line of
sight smaller than $500~\kpc$ relative to the centre of the halo has
been considered. In the leftmost panel, the black circle marks the
position of the halo virial radius ($R_{\rm vir} = 239~\kpc$), whilst
the squares are centred on positions of the main galaxy and of
satellites containing gas. Their sizes are scaled according to the gas mass.}
\label{fig:maps}
\end{figure*}

\begin{enumerate}
 \item primordial and metal cooling rates including self-shielding
   corrections by \citet{Rahmati2013};
 \item a variation of the \citet{SFR_paper} sub-resolution model for the ISM with the adoption of a
       \citet{Chabrier2003} initial mass function;
 \item a self-consistent treatment of stellar evolution, mass return from stars and metal enrichment 
       (tracking H, He, C, N, O, Ne, Mg, Si, Fe and total metallicity) of the gas phase;
 \item galactic winds implemented via a kinetic wind scheme, with mass loading computed according
       to an energy-driven scaling \citep[see][]{Puchwein2013};
 \item wind metal loading offset with respect to the mass loading, necessary to reproduce the
       oxygen abundance of low mass haloes;
 \item BH accretion and feedback based on a modification of the
   \citet{BH_paper} model, accounting for
       quasar, radio and radiative feedback channels;
 \item a spatially uniform UV ionizing background \citep{FaucherGiguere2009};
 \item a new Monte-Carlo formalism for Lagrangian tracer particles \citep{Genel2013}.
\end{enumerate}
We use the same fiducial settings as in \citet{Vogelsberger2013} for
the free parameters of the galaxy formation model, except for two
minor modifications in the wind implementation and the radio-mode
feedback of the BH module, respectively. These modifications are the
same as in \citetalias{Marinacci2013} and consist of providing the
wind with some thermal energy in addition its kinetic energy, and
adopting a more continuous heating model for the radio-mode feedback
(see \citetalias{Marinacci2013} for further discussion).

\section{Results}\label{sec:results}

\subsection{Diffuse gas properties}\label{sec:gas}

Before starting the analysis of the features of the diffuse
circum-galactic gas, we must define how we select this gas component
in the simulations. In the context of our ISM treatment, we consider
as diffuse gas all the gas cells whose density is below the star
formation density threshold ($n_{\rm th}\simeq 0.13~\cm$ for the
fiducial setting adopted in the simulations). We note that this is a
rather broad selection criterion since it encompasses relatively dense
($n\sim 0.1~\cm$) and cold ($T\sim 10^4~\K$) gas in the immediate
proximity and of the main galaxy (and satellites) as well as more
tenuous ($n\sim 10^{-4}~\cm$) and hotter ($T\sim T_{\rm vir}$) gas
phase that is expected to be distributed in the CGM on scales of $\sim
100~\kpc$. The properties of the phases comprising this gas component
are the topic of this section.

\begin{figure*}
\resizebox{17.8cm}{!}{\includegraphics{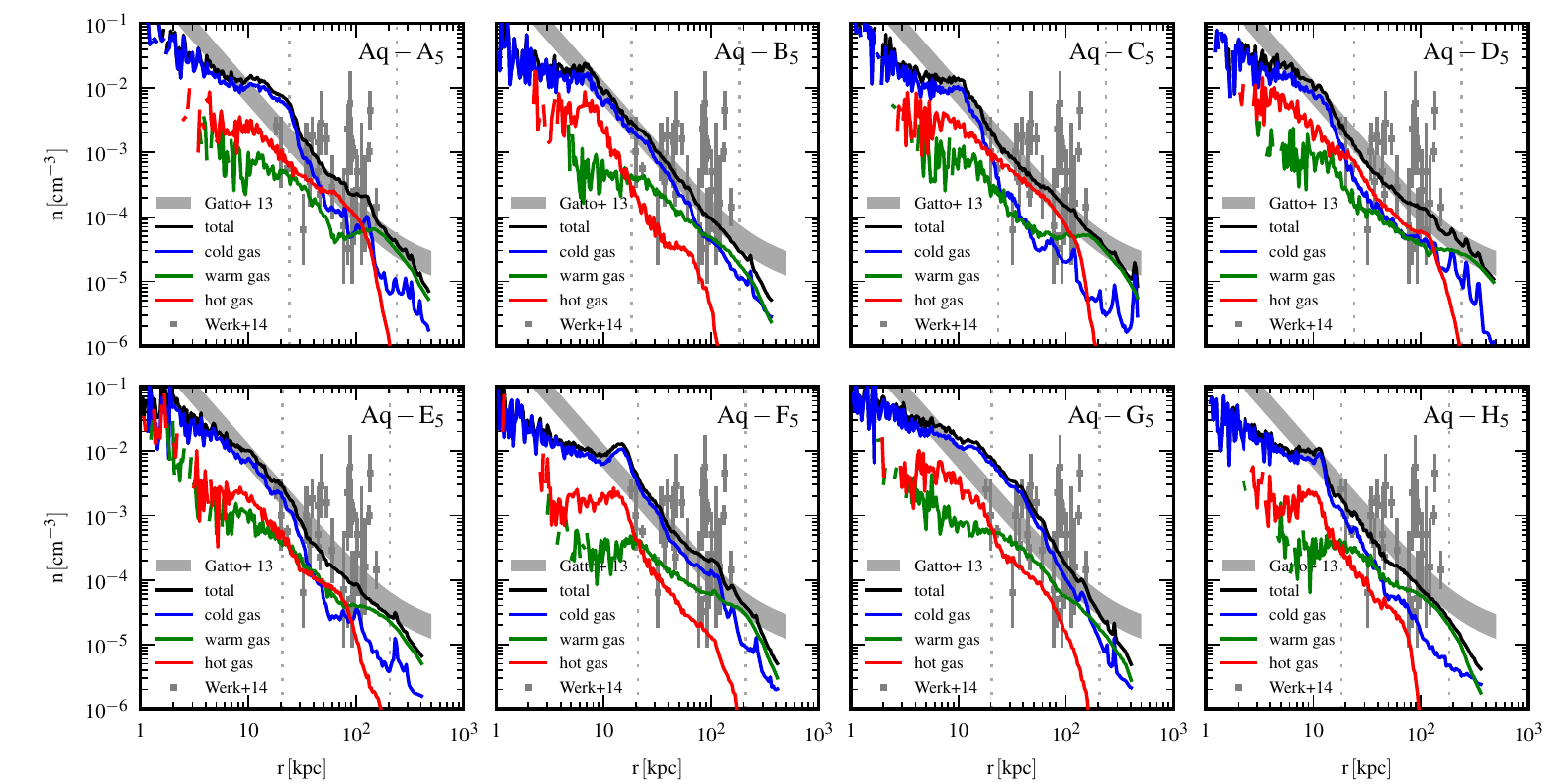}}
\caption{Spherically-averaged density profiles of the diffuse gas as a
function of radius for the whole simulation set at $z = 0$. The
diffuse gas has been divided into a cold phase (blue line) with $T <
10^5~\K$, a warm phase (green line) with $10^5~\K < T < 10^6~\K$ and a
hot phase with $T > 10^6~\K$. The total density profile is indicated by
the solid black line, while the dotted vertical lines are located at
$0.1\times R_{\rm vir}$ and $R_{\rm vir}$, respectively. The gray bands
in all panels show the density range of isothermal 
($T = 1.8\times 10^6\,\K$) models for the Milky Way's corona derived 
from ram pressure stripping simulations by \citet{Gatto2013} that take also
into account X-ray emission and pulsar dispersion measure upper limits 
\citep{Anderson2010}. Gray points with error bars are taken from \citet{Werk2014}
by correcting their hydrogen volume densities for helium and electron
abundances (see text for details).} 
\label{fig:density}
\end{figure*}

In Fig.~\ref{fig:maps}, we present column density maps at $z = 0$ of
the diffuse gas for the case of the simulation Aq-A-5. The total
column density is shown together with the column densities of \hi\ and
oxygen to study both the large-scale morphology and the degree of
chemical enrichment of the CGM. To create the maps, we choose a
projection plane centered on the halo potential minimum, with a size of
$1~\Mpc$ on a side. We project all the gas cells that have a
distance along the line of sight smaller than $500~\kpc$ from the halo
centre.  We thus used a purely spatial selection criterion for the
projections, although it is also possible to associate the diffuse gas to
the halo based on kinematic information \citep{Stinson12}.  The latter
method is what is actually adopted in absorption studies of the CGM
both in the Milky Way \citep[][]{Sembach2003, Savage2003, Lehner2012}
and in external galaxies \citep[e.g.,][]{Prochaska2011} because of the
lack of spatial information along the line of sight about the
absorbers. However, we explicitly checked that our results do not
strongly depend on the way the association between the halo and its
CGM is made. For instance, adopting a cut in systemic velocity
$|\delta v| = 200~\kms$ produces results nearly indistinguishable from
those presented in Fig.~\ref{fig:maps}.  Significant differences
appear only if a rather extreme cut $|\delta v| < 50~\kms$ is used.
In the projections the halo is oriented such that the central galaxy
would be seen edge-on. In other words, this means that the vertical
axis on the maps is aligned with the galaxy symmetry axis, calculated
according to the procedure described in \citetalias{Marinacci2013}. We
choose that particular orientation because it is the most suited to
show the effects of our galactic wind implementation -- which is
launched preferentially perpendicularly to the disc -- on the
structure of the CGM.

The total column density (left panel) of the CGM exhibits a very
regular global morphology and extends in an almost spherical fashion
slightly beyond the virial radius of the halo (the black circle in
Fig.~\ref{fig:maps}), where a more filamentary structure can be seen. Regions of
gas with column densities in excess of $10^{21}~\cmsq$ are usually
compact and associated with the central galaxy and its satellites
(indicated by black squares).  This is also confirmed by
the \hi\ column density map (central panel), that shows indeed that
the vast majority of the high column density gas is accounted for
almost completely by this component. The \hi\ morphology is much less
regular and less extended with respect to the total column
density. The neutral hydrogen tends to be preferentially located close
to density peaks (i.e., in the vicinity of the central galaxy and its
satellites) and drops by several orders of magnitude in between
them. This cold material can be thought of as forming an
interface between the star forming region of a galaxy and its
environment, possibly mediating further cooling of hotter gas and thus
feeding star formation \citep{Marinacci2010, Marinacci2011}.

The spatial distribution of heavy elements is similar for all the
metal species that we track in our galaxy formation implementation. In
the right panel of Fig.~\ref{fig:maps}, we take oxygen as an example.
It can be seen that the distribution of this element is markedly
different from that of the \hi\, and in fact it resembles more closely
that of the total gas. Even in this case oxygen (and the other metals
as well) is present up to and also beyond the virial radius of the
halo. The presence of oxygen in these regions implies that galactic
winds are very effective in ejecting metals from their production
sites, i.e.~the star forming regions of the (central) galaxy, to the
surrounding environment. The fact that galactic winds are the main
agent that causes metal enrichment of the CGM is also indicated by the
morphology of the oxygen enriched region. As mentioned above, it looks
smoother than the \hi\ distribution and similar to the total gas
column density but, compared to the latter, it is more elongated in
the vertical direction, which is the preferred direction of wind
ejection in our implementation. The wind -- which is ultimately
powered by supernovae associated with star formation -- is more
effective in ejecting metal-enriched material in the more active
star-forming regions, such as those corresponding to the central
galaxy, as can be seen from filaments with column density $\sim
10^{17}\cmsq$ emanating from the centre and reaching heights of $\sim
100-150~\kpc$.

\begin{figure*}
\resizebox{17.8cm}{!}{\includegraphics{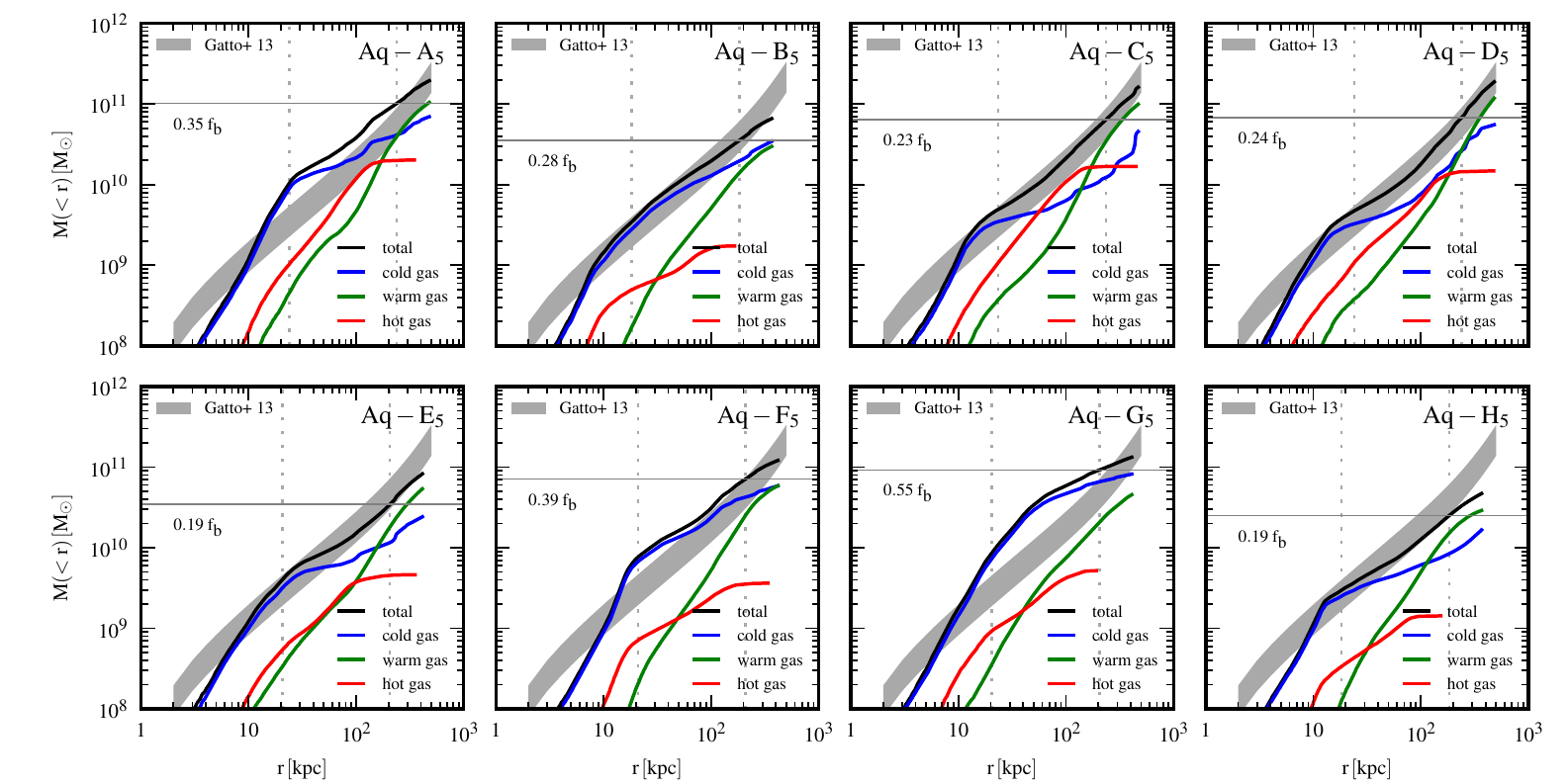}}
\caption{Cumulative mass profiles of the diffuse gas as a function of
  radius for the whole simulation set at $z = 0$. The contributions to
  the total mass of different gas phases, separated according to the
  gas temperature as in Fig.~\ref{fig:density}, are also plotted. The
  thin horizontal line in each panel shows the total mass in diffuse
  gas contained within the virial radius together with the indication
  of how this mass compares to that inferred from the universal baryon
  fraction.  The gray bands in all panels indicate the allowed range
  of cumulative mass profiles of the Milky Way's corona models
  presented in Fig.~\ref{fig:density}.  }
\label{fig:gasmass}
\end{figure*}

In Fig.~\ref{fig:density}, we show spherically-averaged density
profiles of the diffuse gas as a function of radius for the full
simulated set of haloes at $z = 0$. The diffuse gas has been divided
into three phases according to its temperature: a cold phase (blue
lines) with $T < 10^5~\K$, a warm phase (green lines) with $10^5~\K <
T < 10^6~\K$ and a hot phase (red lines) with $T > 10^6~\K$. The total
gas density is indicated by the solid black line. Each profile has
been derived by dividing the radial distance in $100$ equally spaced
logarithmic bins and then computing the gas mass falling into each
bin. The average density then simply follows by dividing the bin mass
by the volume of the spherical shell defined by each radial bin. Each
halo generally features declining density profiles. The decreasing
trend is somewhat less pronounced in the region within $0.1\times
R_{\rm vir}$, which is marked by the first of the two vertical dotted
lines in the Fig.~\ref{fig:density} (the other marking the location of the virial
radius) and roughly coincides with the extension of the star-forming
disc\footnote{We recall that this radial cut was used in
  \citetalias{Marinacci2013} to select the material belonging to the
  central galaxy.}, and then becomes steeper from that point on out to
the virial radius. The density profiles are dominated in the inner
region by the cold phase. This seems to be the general trend also at
larger radii with the noteworthy exceptions of the Aq-A-5 and Aq-C-5
haloes where it is the hot phase that gives the dominant
contribution. Some other haloes (Aq-D-5 and Aq-E-5), instead, show a
roughly equal contribution to the total density of all three phases.

We compare our density profiles to isothermal models for the
Milky Way's corona of \citet{Gatto2013}. These models of are
derived from ram pressure stripping simulations of the dwarf
spheroidal galaxies Sextans and Carina and the allowed density ranges,
shown by the gray bands in Fig.~\ref{fig:density}, take already into
account the observational constraints (not reported in the figure)
coming from X-ray emission and pulsar rotation measures upper limits
in the Milky Way \citep{Anderson2010}. The predictions of our
simulations are consistent with the admissible density ranges of these
models for the Galactic corona. However, we must stress that the match
is excellent for our \textit{total} density profiles, while in the
simulations of \citet{Gatto2013} the corona is assumed to be composed
of hot gas ($T = 1.8\times 10^6\,\K$ in the models shown here) close to
the virial temperature of the Milky Way halo. As we have discussed
above, the contribution of the hot gas to our density profiles is in
general sub-dominant with respect to the cold gas. Only in the Aq-A-5
and Aq-C-5 haloes the density of the hot gas phase agrees with the ram
pressure stripping constraints derived by \citet{Gatto2013}.

We also compare the density profiles obtained in our simulations
to COS-Halos data of low-redshift ($z\sim 0.2$) $L_{*}$ galaxies
\citep[][]{Werk2014} that directly probe the 
condition of the CGM in late-type systems. To get the total density
profiles of the CGM (shown as points with error bars in
Fig.~\ref{fig:density}) we correct the hydrogen density $n_{\h}$ for
helium abundance by assuming primordial composition ($X = 0.76$) and
for electron abundance by adding to the corrected density 1.16$\times
n_{\h}$. The latter value is the mean conversion factor quoted in Fig.
12 of \citet{Werk2014} to obtain their electron density profiles. The
predictions of our simulation for the total density profiles of the
CGM are in agreement with these observational findings.

\begin{figure*}
\resizebox{17.8cm}{!}{\includegraphics{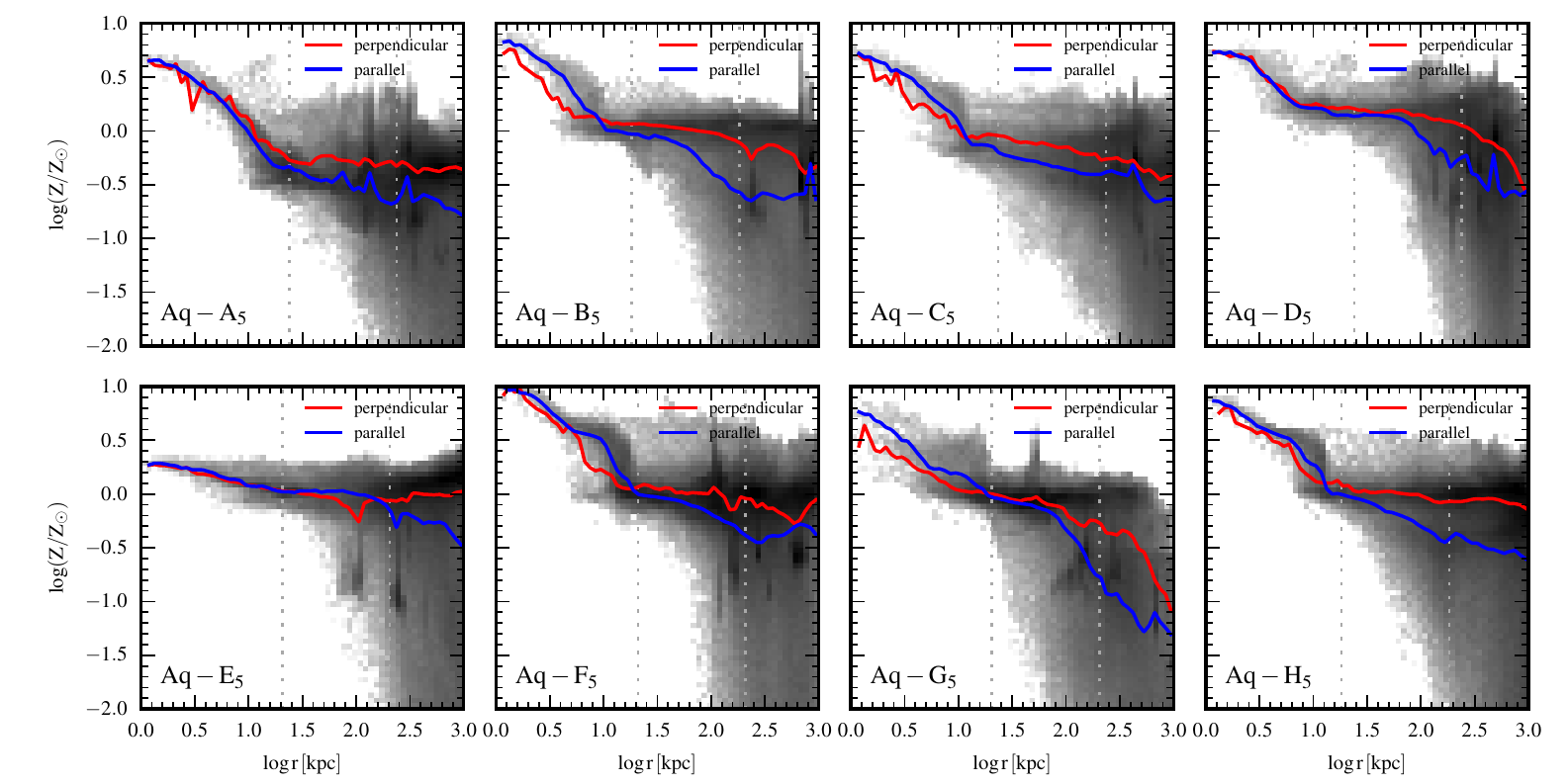}}
\caption{Metallicity distribution versus radial position of the
  diffuse gas phase for all the simulated haloes at $z = 0$. The
  figure shows a two-dimensional histogram built on $60\times60$
  equally spaced logarithmic bins and weighted by the mass of gas
  cells so that the gray-scale encodes the gas mass present in each
  bin (darker colours indicate larger masses). The red solid lines in
  each panel represent the (mass-weighted) average gas metallicity as
  a function of the radius in a cone centered on the symmetry axis of 
  the galaxy with an opening angle of 30 degrees, while the solid blue 
  lines show the same quantity in the region complementary to that cone.
  }  
\label{fig:metalposhist}
\end{figure*}

Being the dominant phase in terms of density, it is not surprising
that most of the mass of the diffuse gas in the simulations is
contained in the cold phase. This is shown in Fig.~\ref{fig:gasmass},
which presents cumulative mass profiles for the three phases defined
above. The profiles are computed by ordering the gas cells according
to their distance from the halo centre and summing their masses up to
a given radius $r$. The thin horizontal line in each panel represents
the total mass in diffuse gas inside the virial radius, and for each
system we also report how this quantity compares to the total baryonic
mass of the halo as inferred from the universal baryon
fraction. Usually, the mass fraction contained in the diffuse gas is
not negligible, ranging from $\sim20\%$ to $\sim40\%$ of the total
baryonic mass in the halo predicted by cosmology, but not dominant
either, since only in the case of the Aq-G-5 halo it accounts for about
the $55\%$ of the total mass. As mentioned above, most of the mass is
contained in the cold phase that shows a rapidly growing profile in
the inner $20~\kpc$ (which approximately corresponds to $0.1\times
R_{\rm vir}$ of the haloes) with a tendency of becoming flatter in the
outer regions out to the virial radius. This flattening is always
present but can sometimes be more pronounced depending on the relative
contribution to the total mass of the warm and hot phases; the most
extreme case being the Aq-C-5 halo. For what concerns the (cumulative)
mass profiles of the warm and hot phases they present the same trend
in the inner regions of the cold counterpart (i.e., a rapid growth)
but the mass of the warm gas continues to steadily increase up to the
halo virial radius, while the mass of hot gas shows a flattening that
is also found in the cold phase. The flatter trend starts at
approximately the same radius as that of the cold gas in half of the
haloes (Aq-B-5, Aq-F-5, Aq-G-5 and Aq-H-5), while in the others this occurs
beyond $100~\kpc$. 

A comparison of our results with those of \citet{Gatto2013} yields
again an excellent agreement with respect to the total mass profiles,
with values of the enclosed masses within the virial radius that are
fairly similar in both cases. It is very encouraging that the two
types of simulations reach consistent conclusions for the properties
of the circum-galactic gas although they study different phenomena and
also differ in the adopted numerical methodology. We stress again
that this agreement holds for the \textit{total} mass profile in our
simulations, while the \citet{Gatto2013} models describe the hot
($T\sim T_{\rm vir}$) phase of the circum-galactic gas. Our total CGM
masses also provide a good match to the masses inferred by
\citet{Werk2014} although their fiducial model predicts that the CGM
accounts for $\sim 45\%$ of the total baryonic mass associated with the
$\sim 10^{12}~M_{\odot}$ dark matter halo of a typical $L_{*}$ galaxy,
a value close to the upper range of what we find in our simulation
set. However, we want to note that CGM mass determinations are very
challenging to obtain observationally and highly uncertain. Depending
on the assumptions made (especially for what concerns the warm and hot
phases of the diffuse gas) the contribution of the CGM to the total
baryon fraction can vary by more than a factor of 2 \citep[see Fig. 11
of][]{Werk2014}.

\begin{table*}
\begin{tabular}{lcccccc}
\hline
Simulation & $r_{\rm max}$ & $M_{Z}(< r_{\rm max})$ & $f_{Z}(< 0.1\times R_{\rm vir})$  &
             $f_{Z}(< 150~{\rm kpc})$ & $f_{Z}(< R_{\rm vir})$ & $f_{b}(< R_{\rm vir})$ \\
           & $(R_{\rm vir})$ & $(10^{10} M_{\odot})$ & & & & \\
\hline
Aq-A-5 & 1.0 & 0.102 &   0.514&     0.844&     1.000  &      \\
       & 2.5 & 0.161 &   0.328&     0.538&     0.637  & 0.55 \\            
       & 5.0 & 0.200 &   0.264&     0.433&     0.512  &      \\
Aq-B-5 & 1.0 & 0.111 &   0.686&     0.956&     1.000  &      \\
       & 2.5 & 0.139 &   0.548&     0.764&     0.800  & 0.70 \\        
       & 5.0 & 0.180 &   0.425&     0.593&     0.621  &      \\            
Aq-C-5 & 1.0 & 0.126 &   0.665&     0.863&     1.000  &      \\           
       & 2.5 & 0.218 &   0.386&     0.501&     0.580  & 0.51 \\
       & 5.0 & 0.258 &   0.326&     0.423&     0.490  &      \\           
Aq-D-5 & 1.0 & 0.225 &   0.635&     0.835&     1.000  &      \\ 
       & 2.5 & 0.372 &   0.385&     0.506&     0.606  & 0.68 \\
       & 5.0 & 0.406 &   0.353&     0.464&     0.556  &      \\
Aq-E-5 & 1.0 & 0.098 &   0.637&     0.880&     1.000  &      \\
       & 2.5 & 0.161 &   0.390&     0.538&     0.611  & 0.67 \\
       & 5.0 & 0.246 &   0.255&     0.351&     0.399  &      \\
Aq-F-5 & 1.0 & 0.227 &   0.615&     0.922&     1.000  &      \\
       & 2.5 & 0.276 &   0.507&     0.759&     0.824  & 0.96 \\
       & 5.0 & 0.311 &   0.449&     0.673&     0.730  &      \\
Aq-G-5 & 1.0 & 0.224 &   0.513&     0.960&     1.000  &      \\
       & 2.5 & 0.244 &   0.472&     0.883&     0.920  & 1.03 \\
       & 5.0 & 0.247 &   0.466&     0.872&     0.908  &      \\
Aq-H-5 & 1.0 & 0.116 &   0.807&     0.964&     1.000  &      \\
       & 2.5 & 0.143 &   0.657&     0.785&     0.814  & 0.61 \\
       & 5.0 & 0.182 &   0.514&     0.614&     0.636  &      \\
\hline
\end{tabular}
\caption{Total metal masses (i.e., including stars and
star-forming gas) and metal fractions for all the simulated haloes at
$z = 0$. The columns give (from left to right): simulation name;
reference radius (in units of the halo virial radius) within which
total metal masses are computed; total metal masses (in units of
$10^{10}~M_{\odot}$); fractions of metals within $0.1\times R_{\rm
vir}$, $150~\kpc$ and $R_{\rm vir}$; and baryon fractions within
$R_{\rm vir}$ relative to the cosmological mean.}
\label{tab:metals}
\end{table*}

To assess the degree of metal enrichment of the diffuse gas phase as a
function of the distance from the central galaxy, we display in
Fig.~\ref{fig:metalposhist} two-dimensional histograms obtained by
sampling the $\log~Z-\log~r$ plane in $60\times 60$ equally spaced
bins for all the simulated haloes. The gray-scale encodes the gas mass
present in each bin, with darker gray shades indicating larger values
of the gas masses. We also show with solid lines two average
mass-weighted metallicity profiles obtained by considering two
different regions of space surrounding the main galaxy: a cone centred
on the galaxy's symmetry axis (and therefore perpendicular to the
star-forming disc) with an opening angle of 30 degrees (red) and the
region complementary to that cone (blue). In all the haloes the
diffuse gas phase is substantially metal enriched. Inside the central
galaxy (i.e. for radii below $0.1\times R_{\rm vir}$) maximum
metallicities of $3$ up to $\sim 10$ times the solar value are not
uncommon. Outside the galaxy, the average gas metallicity shows in
general a rather flat trend with radius that in some cases becomes
steeper in the proximity of the virial radius. Beyond $0.1\times
R_{\rm vir}$ the average metallicity in the cone perpendicular to the
disc is close to the solar value. The only exceptions are the Aq-A-5,
Aq-C-5 and Aq-G-5 systems, which feature slightly sub-solar
metallicities. The metallicity profiles in the region outside the cone
have a somewhat lower value of the metallicity with respect to the
region inside the cone and only in the case of the Aq-E-5 halo the two
profiles agree. This trend is either inverted or the two profiles have
essentially the same metallicity values inside the central galaxy.

The presence of a significant degree of metal enrichment in the haloes
of our simulated galaxies demonstrates again the effectiveness of
galactic winds in transferring metals from their production sites (the
star-forming discs) to the circum-galactic regions surrounding them.
Moreover, the scatter in the metallicity distribution tends to
increase with radius. This implies that at any given radius, regions
with rather different metallicity co-exist or, in other words, that
metals are not homogeneously mixed in the gas. The bipolar nature of
our galactic wind model (see also the right panel of
Fig.~\ref{fig:maps}) spontaneously leads to this type of behaviour
because of its intrinsic anisotropy, which is also the cause for
the differences between the metallicity profiles perpendicular and
parallel to the disc discussed above.

To further characterize the degree of metal enrichment of the CGM,
in Table~\ref{tab:metals} we present the total metal masses and metal
fractions for all the simulated haloes at $z=0$. Note that all the
entries in the table include also the contributions of stars and
star-forming gas to properly account for all the metal content of the
simulated haloes. These components, however, only contribute to the metal 
content of the central galaxy, where they are located. We list the total 
metal masses within three
different fiducial radii (namely 1.0, 2.5 and 5.0 times $R_{\rm vir}$)
and the corresponding metal fractions within 0.1 $R_{\rm vir}$, 150
kpc \citep[the reference radius in][]{Peeples2014} and $R_{\rm vir}$.
Finally, in the last column we report the baryon fractions within the
virial radius taken from \citetalias{Marinacci2013}.

Metal masses steadily increase as the fiducial radius increases,
implying that a non-negligible fraction metals is located beyond the
virial radius of all the simulated systems. For a reference radius of
$2.5\times R_{\rm vir}$ this fraction can be as high as $40\%$ of the
total metal mass. At the same fiducial radius, the metal fractions
within 150 kpc are in the range $\sim 50-80\%$ of the total metal
masses. While the lower value of this range it is still consistent
with that presented by \citet{Peeples2014}, who report that for a
typical $L_{*}$ galaxy only $\sim 40\%$ of the total metal mass should
be found inside $150~\kpc$, about half of our objects have metal
fractions close to $80\%$. Interestingly, higher baryon fractions
appear to correspond to higher metal fractions, although this
correlation is not particularly strong.  Most of the metals in the
haloes are found within $0.1\times R_{\rm vir}$ -- the adopted size of
the central galaxy -- and are contained in stars and star-forming gas.
This component accounts for more than 50\% of the metal mass contained
within 150 kpc for all the haloes.

\begin{figure*}
\resizebox{17.8cm}{!}{\includegraphics{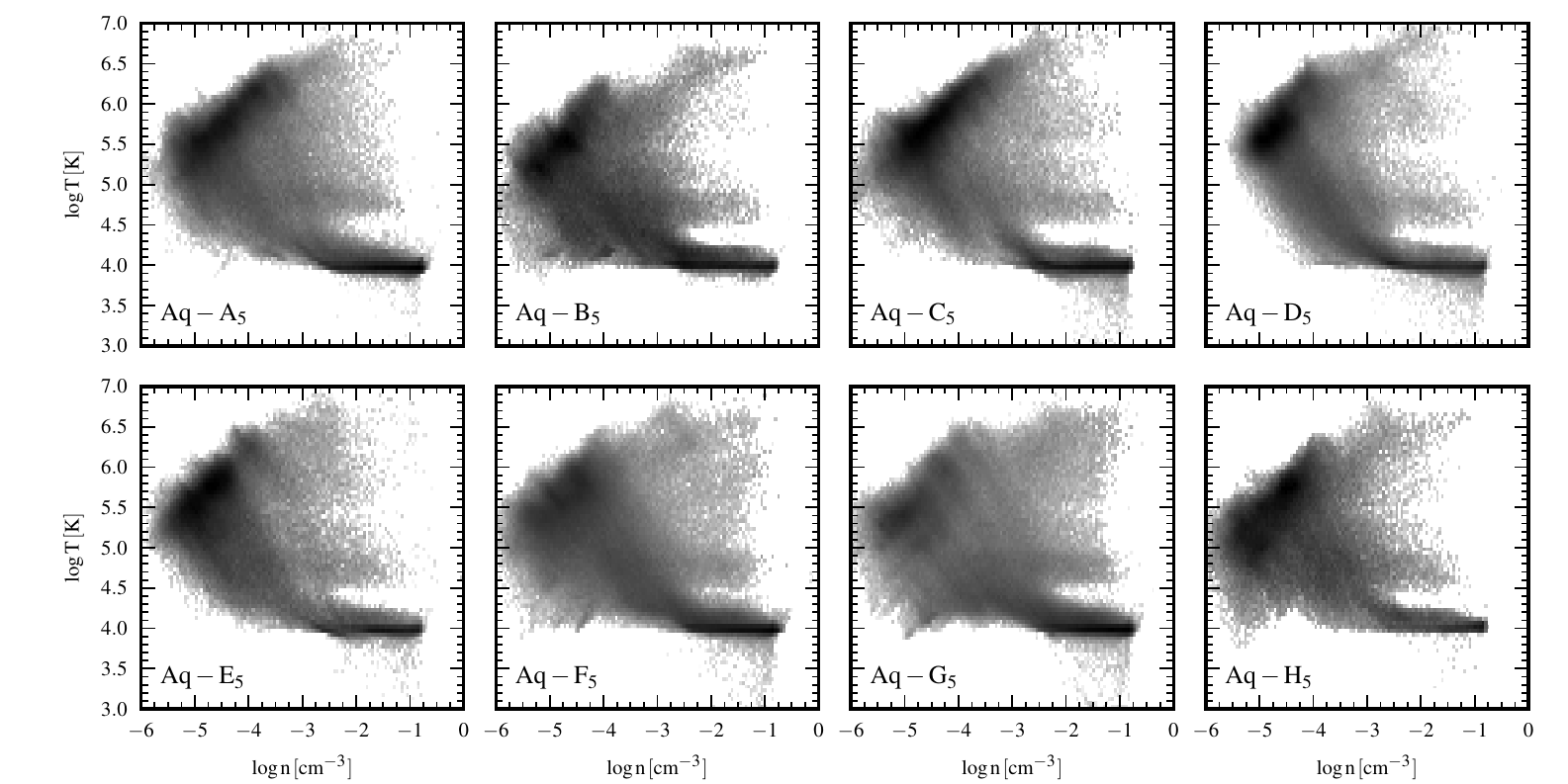}}
\caption{Phase diagram of the diffuse gas phase for all the simulated
  haloes at $z = 0$. The $\log~T-\log~n$ plane is divided in
  $80\times120$ equally spaced bins on which a two-dimensional
  histogram is built. The histogram is weighted by the gas mass in
  each cell and the gray-scale encodes the amount of mass present in
  each bin (darker colours indicate larger values of the mass). It is
  readily apparent that the diffuse medium tends to concentrate in two
  distinct parts of these diagrams, one located towards high densities
  ($10^{-2}\cm < n \lsim 1~\cm$) and low temperatures ($T \sim 10^{4}\K$),
  and the other at low densities ($10^{-5}\cm < n < 10^{-3}\cm$) and
  temperatures close to the virial one ($\sim10^6~\K$).}
\label{fig:phasediagram}
\end{figure*}

In Fig.~\ref{fig:phasediagram} we present the phase diagram of the
diffuse gas for all the simulated haloes at $z = 0$. Specifically, we
built a two-dimensional histogram by dividing the $\log~T-\log~n$
plane in $80\times 120$ equally spaced bins and plotting, as a
gray-scale, the total gas mass inside each bin (darker gray shades are
assigned to larger values of the mass). From
Fig.~\ref{fig:phasediagram}, it is clear that the diffuse medium it is
not uniformly distributed in the $\log~T-\log~n$ plane but gathers in
two definite regions of the diagrams. One region is located at low
densities ($10^{-5}\cm < n < 10^{-3}\cm$) and at temperatures $\log~T
\gsim 5.5$ and centred on the halo virial temperature, which for the halo
mass range that have been explored in these simulations is $\sim
10^6~\K$. This warm-hot gas may be identified with the extended
component of the CGM seen in the left and right panels of
Fig.~\ref{fig:maps}. The other region is at higher densities
($10^{-2}\cm < n \lsim 1~\cm$) and at an almost constant temperature of
$10^4~\K$.  This cold and relatively dense gas is closely related to
the \hi\ map presented in the central panel of Fig.~\ref{fig:maps}
from which it can again be seen that this gas is found in the
proximity of the central galaxy and its satellites and therefore has a
less extended spatial distribution with respect to the warm-hot
component. The transition between these two gas phases is not sharp
but rather continuous. Indeed, the remaining part of the diagram in
between the warm-hot and the cold dense gas is populated by an
intermediate gas phase that can either result from the cooling of the
denser and hotter fraction of the warm-hot phase via radiative losses
-- the less dense fraction can have a cooling time comparable or
longer than a Hubble time -- or the heating of the cold phase by the
thermalization of the galactic wind energy or AGN feedback.

\begin{figure*}
\resizebox{17.8cm}{!}{\includegraphics{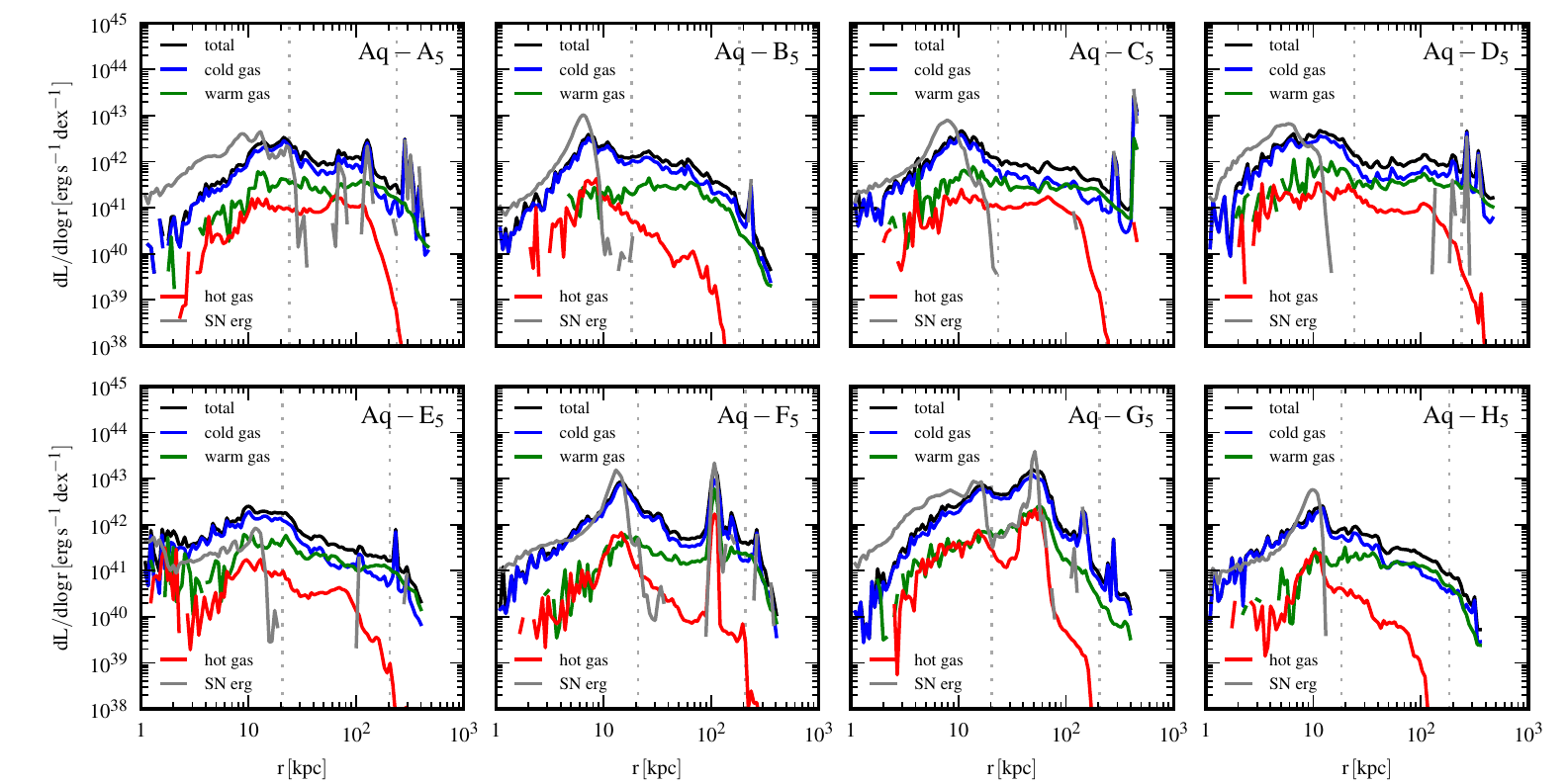}}
\caption{Spherically-averaged differential (bolometric) luminosity
  profiles for the diffuse gas phase for all the simulated haloes at
  $z = 0$, computed as the total cooling rate in equally
  logarithmically spaced radial bins divided by the bin width. The
  relative contributions of the cold, warm and hot components,
  selected according to the the gas temperature (see also
  Figs~\ref{fig:density} and \ref{fig:gasmass} above), are shown
  together with the energy of supernovae (gray lines) that is
  contained in galactic winds and ultimately deposited in the CGM of
  the simulated galaxies. Peaks in the latter quantity are usually 
  found at the same locations of those of the total gas luminosity,
  suggesting that the energy carried by galactic winds is one of the
  major factors which sets the thermal state of the diffuse gas.  }
\label{fig:diffluminosity}
\end{figure*}

To quantify how the energy injected by the wind into the CGM affects
the thermodynamic state of the latter, we show in
Fig.~\ref{fig:diffluminosity} the spherically-averaged differential
luminosity profiles of the diffuse gas together with the energy
produced by stellar feedback that is available for galactic
winds. Again, we split the contribution to the total gas luminosity
between its cold, warm and hot phases. We derived the luminosity
profiles by computing the total cooling rate of the gas (and thus the
luminosities that we plot are \textit{bolometric} luminosities) in
$35$ equally spaced logarithmic radial bins and then dividing the
resulting rate by the bin width. The energy available to galactic
winds is estimated by adding up the star formation rate (SFR) of
star-forming gas cells within a given bin. This quantity is then 
converted into an energy rate by using a factor that expresses
the supernova energy per unit mass for a \citet{Chabrier2003} initial
mass function times the assumed wind efficiency factor \citep[for
details see][section 2.5.1]{Vogelsberger2013}. Also in this case the
radial profile is obtained by dividing the supernova energy rate thus
computed by the bin width.

Several important aspects can be appreciated from
Fig.~\ref{fig:diffluminosity}. Perhaps the most important one is that
the energy in the galactic winds, that in our model contain nearly
all the energy originating from supernovae, can easily compensate for
the radiative losses of the gas. This immediately suggests that
the wind energy is one of the major energy sources of the diffuse
medium, if not the most prominent. Most of the wind energy is
generated within the inner $\sim 20~\kpc$ of the haloes (roughly the
size of the star-forming region), where the central galaxy is located
and star formation is more vigorous. In this region, the supernova
energy available to winds \FM{is comparable and in several cases even offsets} 
the radiative losses of the gas, the only exception being the Aq-E-5 halo. 
Indeed, this object has
the lowest SFR at redshift $z=0$ of the whole simulation set
\citepalias[see Fig. 14 of ][]{Marinacci2013}, which implies a lower
supernova rate and hence less energy available to generate galactic
winds. The energy produced in the central regions is transferred to the
CGM through the interaction between galactic winds and the CGM itself.
This interaction eventually leads to the thermalization of the wind
energy, which is thus available for powering the emission of the
circum-galactic gas.
The supernova rate suddenly drops at radii greater than $\sim 20~\kpc$
and so obviously does the wind energy, although occasional peaks of
this quantity, very likely due to the presence of star-forming
satellites, can be observed in the external regions. The presence of
these peaks is very interesting because they are always accompanied by
a corresponding peak in the gas luminosity. This further
reinforces the idea that galactic winds are an important energy input
sustaining the gas radiative losses.

\begin{figure*}
\resizebox{17.8cm}{!}{\includegraphics{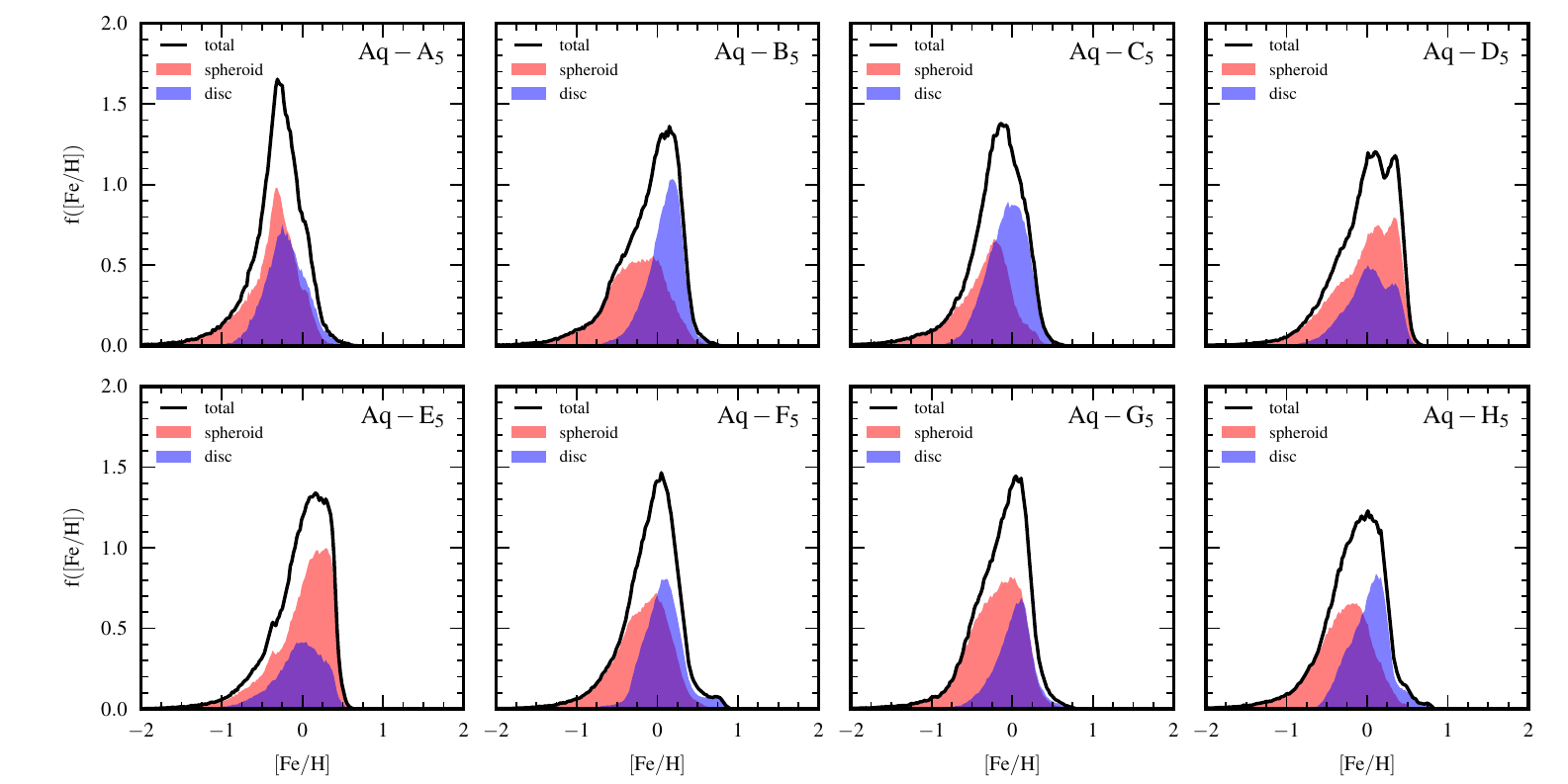}}
\caption{Histograms of the iron abundance relative to the solar value
  of stars within $0.1\times R_{\rm vir}$ for all the simulated haloes
  at $z = 0$. The contributions of disc and bulge stars, separated
  according to their circularity parameter $\epsilon$, are also
  shown. The histograms are normalized such that the area below them
  represents the stellar mass fraction in each component. Most of the
  histograms present a distribution of stellar metallicities peaked at
  around $[{\rm Fe/H}] \sim 0$ and skewed towards sub-solar
  values. This general behaviour results from the superposition of the
  contributions of disc stars that are on average more metal-rich
  (because they are younger) and more narrowly distributed around the
  location of the peak of their metallicity distribution, and less
  metal-rich bulge stars that have a broader distribution with a tail
  towards lower metallicities.}
\label{fig:starFehistogram}
\end{figure*}

Another key point is the fact that almost all of the gas luminosity is
accounted for by the cold phase up to distances of $\sim 100~\kpc$,
where usually the warm phase gives a similar contribution. The hot
phase is always sub-dominant in the luminosity budget. This is not
surprising because gas in the cold and warm phases have temperatures
that encompass the range where cooling is enhanced by the presence of
metals (for solar metallicity, the cooling function peaks at a few
$\times 10^5~\K$). Furthermore, in the cold phase the enhancement due
to the presence of metals is further assisted by a greater gas
density. The hot phase lacks both of these boost factors: it has
sufficiently high temperature to be located away from the peak of the
cooling function -- and actually at temperatures of $\sim 10^6~\K$ it
is close to a \textit{local minimum} of the cooling function before
the cooling rate starts to increase again due to free-free emission --
and its density is low, rarely exceeding a few $10^{-3}~\cm$ and only
in innermost regions (see Fig.~\ref{fig:density}).

\subsection{Stellar metallicities} \label{sec:stars}

We now pass to analyse the metal content of stars comprising the
central galaxy in our simulations to see how this is affected by
strong galactic outflows. We first focus on the global metal content
(Sec.~\ref{sec:globalmet}) and we then discuss how the metal
abundance varies as a function of the galactocentric radius
(Sec.~\ref{sec:gradients}).

\subsubsection{Global metallicity distribution} \label{sec:globalmet}

We start our analysis of stellar metallicity by presenting first its
global distribution. We will focus on stars belonging to the central
galaxy adopting our usual spatial cut that includes only stars inside
$0.1\times R_{\rm vir}$ of the host halo. Since the virial radii of
all the simulated objects are $\approx 200~\kpc$, this spatial cut
will include disc and bulge stars, as well as so-called halo stars
(i.e., stars outside the disc of the central galaxy and
gravitationally bound to it).

In Fig.~\ref{fig:starFehistogram}, we show histograms of the iron
distribution -- a widely used proxy for the total metal content --
relative to the solar value \citep[taken from table 5 of][]{Asplund09}
of such stars for all the simulated objects at $z = 0$. The histograms
are normalised to the total stellar mass, and the contributions of
disc (blue area) and bulge (red area) stars are also displayed. To
separate these contributions we adopted the same kinematic criterion
as in \citetalias{Marinacci2013} based on the circularity parameter
$\epsilon$ of stellar orbits, so that stars with $\epsilon > 0.7$ are
considered as being part of a rotationally supported disc while the
remaining part comprises the bulge. The total iron distribution has a
peak at ${\rm[Fe/H]\sim 0}$ in almost all the haloes. The only
exception is Aq-A-5, where the peak of the distribution is shifted
towards lower metallicities, giving ${\rm[Fe/H] \sim -0.8}$. Usually,
the iron distributions are unimodal and rather symmetric around their
maximum value, although with a tail at sub-solar
metallicities. Occasionally a second peak, as in the case of the Aq-D-5
halo, or a small bump at super-solar metallicities, for instance in
the Aq-F-5 and Aq-H-5 haloes, may be present.

\begin{figure*}
\resizebox{17.8cm}{!}{\includegraphics{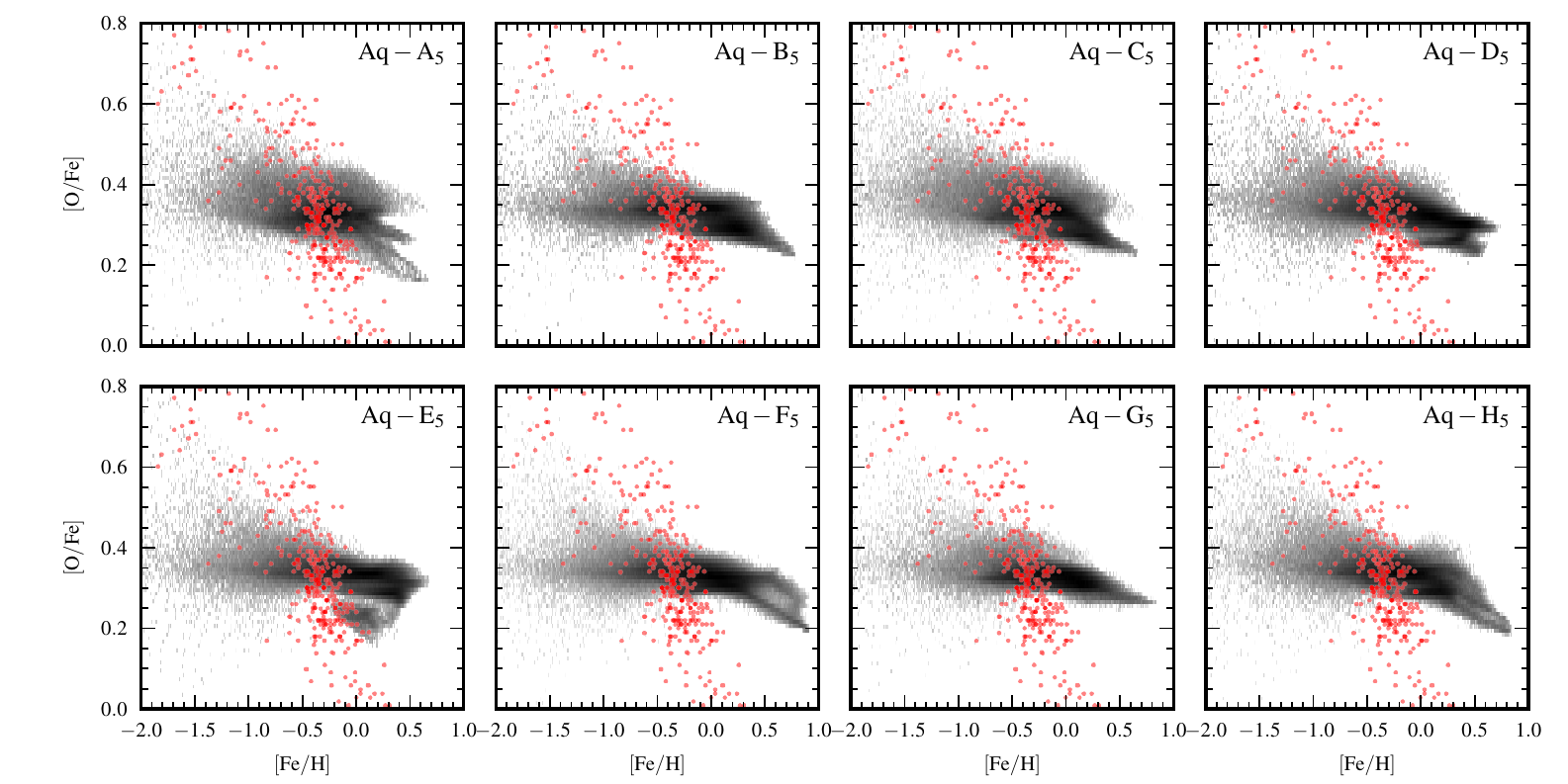}}
\caption{Oxygen versus iron abundance of stars within $0.1\times
  R_{\rm vir}$ for all the simulated haloes at $z = 0$. The $[{\rm
    O/Fe}]-[{\rm Fe/H}]$ plane is divided in $80\times300$ equally
  spaced bins on which a two-dimensional histogram is built. The
  histogram is weighted by the star particle mass and the gray-scale
  encodes the stellar mass present in each bin (darker colours
  indicate larger masses). In the $[{\rm O/Fe}]-[{\rm Fe/H}]$ plane
  stars form a sequence that for increasing values of the iron
  abundance shows a slowly decreasing trend in the $[{\rm O/Fe}]$
  values, although a substantial scatter is present. 
  However, when compared to Milky Way data (red
  points) the trend is much flatter than the observed one. Especially at
  $[{\rm Fe/H}]\sim 0$ our results are unable to reproduce the sharp
  cut-off in the $[{\rm O/Fe}]$ abundances present in the data.  }
\label{fig:oxygenvsironstars}
\end{figure*}

The features of the total metallicity distributions are the result of
the superposition of the distributions of disc and bulge stars that
have quite different properties. Both distributions are unimodal
(except again those of Aq-D-5) but they usually peak at different
metallicities. For disc stars the peak is located at solar or slightly
super-solar metallicities, while bulge stars are less metal-enriched
and the peak of their distributions is located at sub-solar
values. This is not surprising because the iron abundance (total
metallicity) is a good tracer of the stellar age. Since disc stars are
on average younger than bulge stars in this set of simulations
\citepalias[see][]{Marinacci2013}, they form from a more
metal-enriched interstellar medium, while older bulge stars are
composed from more pristine material. However, not all the haloes
feature peaks at two different locations. In particular, there are two
exceptions, namely haloes Aq-D-5 and Aq-E-5, where the metallicities of
disc and bulge stars are very similar. It is interesting to notice
that these two exceptions are also the two most bulge-dominated
galaxies of our whole simulation set. Another distinctive difference
between the metallicity histograms of disc and bulge stars is that the
former are more narrowly distributed around their peak value, whilst
the latter have a broader distribution and a significant tail at lower
metallicities which is absent for disc stars.  Due to the narrowness
of the distribution, and also to the fact that most of these
systems are disc-dominated, the peak of the distribution is usually
more prominent for disc stars.

In Fig.~\ref{fig:oxygenvsironstars}, we present two-dimensional
histograms of the ${\rm [O/Fe]}$ abundance versus the ${\rm [Fe/H]}$
abundance at $z = 0$ for all the simulated haloes. The histograms are
generated by dividing the ${\rm [O/Fe]}-{\rm [Fe/H]}$ plane into
$80\times 300$ equally spaced bins and computing the stellar mass
falling into each bin. This quantity is encoded by the gray scale in
Fig.~\ref{fig:oxygenvsironstars}, with darker gray shades assigned to
larger masses.  From stellar evolution theory an anti-correlation
between the two abundances is expected. The strength of the
anti-correlation (i.e., the slope in the diagram) is of course
influenced by the details of the star formation history of the galaxy,
but its origin lies mainly in the channels that are responsible for
the production of oxygen (and the so-called $\alpha$ elements of which
oxygen is the most abundant) and iron. Oxygen is mostly produced in
core-collapse supernovae generated by the explosion of massive stars
($> 8~\mo$), while the main channel for iron production is represented
by type Ia supernovae -- thermonuclear explosions of white dwarfs.
Because of the mass difference in the progenitors of these classes of
objects, their explosion times are very different. In particular,
core-collapse supernovae explode promptly, enriching the ISM rapidly
with $\alpha$ elements, while type Ia supernovae need more time before
they can start the iron enrichment. This implies that younger stars,
that form from a more type Ia-enriched ISM and thus contain more iron,
will have on average a small ${\rm [O/Fe]}$ ratio since there has been
enough time for type Ia supernovae to efficiently enhance the
abundance of iron with respect to the $\alpha$ elements. Given its
origin, reproducing this anti-correlation is therefore an important
test for the stellar evolution model and metal enrichment scheme
employed in our simulation set.

\begin{figure*}
\resizebox{17.8cm}{!}{\includegraphics{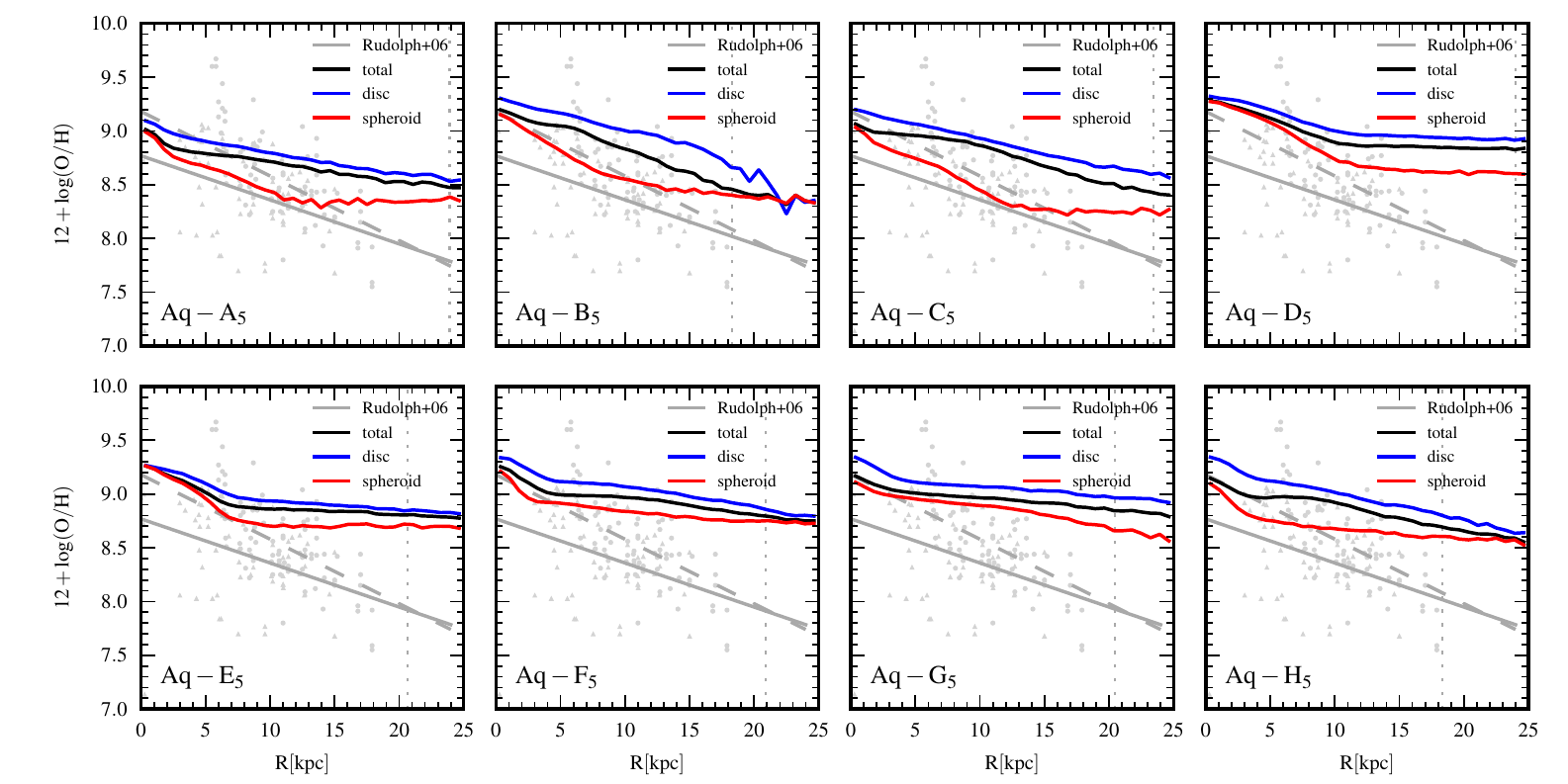}}
\caption{Stellar oxygen abundance as a function of galactocentric
  radius of star particles with $R < 25~\kpc$ and $|z| < 0.1\times
  R_{\rm vir}$ for all the simulated haloes at $z = 0$. The plots show
  the radial abundance profiles in three different cases: the total
  sample of stars (black lines), the disc stars (blue lines) and bulge
  stars (red lines). Gray points are Milky Way data taken from
  \citet{Rudolph06} with gray lines representing two different fits of
  the data in the case of IR derived metallicities (solid lines) or
  optical derived metallicities (long-dashed lines). The vertical
  dotted lines are located at $0.1\times R_{\rm vir}$. The oxygen
  abundances predicted by the simulations are in agreement with the
  observed values and a metallicity gradient is clearly visible in all
  the panels for all the considered components, although it is always
  too shallow when compared to the observations.}
\label{fig:oxygenprofilestar}
\end{figure*}

From Fig.~\ref{fig:oxygenvsironstars} it can be seen that 
at low metallicity (low values of ${\rm
  [Fe/H]}$) there is significant scatter in the histograms, but this
is reduced at higher metallicities (${\rm [Fe/H]} > -0.5$) where a
well-defined sequence forms in all the panels. Most of the times the
sequence is not unimodal but there is a hint of a gap located at ${\rm [O/Fe]}
\sim 0.3$. The declining trend of ${\rm [O/Fe]}$ with ${\rm [Fe/H]}$
is present in all the panels, but it is never very pronounced and
tends to steepen, sometimes considerably, with increasing metallicity.
Interestingly, in two of the haloes that show a double sequence,
namely the Aq-B-5 and the Aq-C-5 haloes, the sequence with the higher
value of ${\rm [O/Fe]}$ seems to show an inversion of the relation. A
comparison with Milky Way data (red points) for halo, thick disc and
thin disc stars\footnote{Observational data is a compilation from
  \citet{Gratton2003}, \citet{Reddy2003},
  \citet{Bensby2005} and \citet{Ramya2012}.} reveals, however, that
the observed anti-correlation is much steeper than that present in our
simulations. In particular, the oxygen abundance predicted by the
simulations matches the observed values only in a narrow range of iron
abundance between $-0.5$ and $0$. At lower metallicities, some
isolated histogram bins can reproduce the observed $[{\rm O/Fe}]$
ratios of the most metal-poor stars -- we recall that data points are
individual stellar measurements -- but the bulk of the distribution
has an oxygen abundance $\sim0.3~{\rm dex}$ below the observed values.
The disagreement is about a factor of 2 and it may just be an
indication of the uncertainties in the elemental yields that we
adopted for low metallicity stars. We also note that, as far as the
lower metallicity part of the panels is concerned, other simulations
reach similar results \citep[see e.g.][]{Gibson2013}. At higher
metallicities, the discrepancy is more marked and our results are not
able to capture the sharp cut-off in oxygen abundance observed at
$[{\rm Fe/H}] \sim 0$. The likely cause of this behaviour is either a
slight over-production of oxygen in our simulations or an
underproduction of iron due to a too low SNIa rate at late times,
a question we further examine in Sec.~\ref{sec:discussion}.

\subsubsection{Radial metallicity gradients} \label{sec:gradients}

Since in the simulations the metallicity information of stars is
available together with their positions, it is interesting to
investigate how the (average) metallicity of the stellar component
varies as a function of position. In particular, several observational
surveys of the Milky Way disc \citep[see][and references
therein]{Rudolph06, RAVEMetals} and of external spiral galaxies
\citep[e.g.,][]{Sanders2010, Barker2007, Magrini2009} report that the
amount of metals is a decreasing function of the galactocentric
distance of stars. It is therefore important to compare the
predictions of our simulations with this observational fact.

We do this for the oxygen abundance in
Fig.~\ref{fig:oxygenprofilestar}, where we present the average
abundance profile of oxygen as a function of the galactocentric radius
$R$ for all the simulated haloes at $z = 0$. The abundance profile is
derived by subdividing the radial range in $35$ equally spaced bins
and then by computing the total oxygen mass of the stars within a
given bin divided by the total stellar mass in that bin. To use the
same radial extent in all the panels, we have somewhat modified the
spatial selection criterion for the stars, which are now included in
the calculation only if they are inside a cylinder of radius $25~\kpc$
that extends above and below the plane of the galactic disc for
$0.1\times R_{\rm vir}$. This distance is indicated in all the panels
by the vertical gray dotted line. We also show, as gray points, the
observed metallicities of 117 Galactic \hii\ regions taken from
\citet{Rudolph06} together with two best-fitting relations derived by
considering the sub-sample of sources whose metallicities are
estimated through IR observations (solid gray line) or optical
observations (long-dashed gray line). We recall that \hii\ regions are
kept ionised by the radiation of very young stars located at their
centres, and therefore their metallicity is a reliable indicator of
the metal content of the stars at their interior. In
Fig.~\ref{fig:oxygenprofilestar}, we separately present the oxygen
profiles for the total sample (black line), the disc (blue), and the
bulge (red) stars. The separation in bulge and disc stars is performed
based on the same criterion adopted in section \ref{sec:globalmet}.

By inspecting Fig.~\ref{fig:oxygenprofilestar}, one can immediately
see that the oxygen abundance generally declines as a function of
galactocentric radius in all the haloes. In agreement with the
findings of Fig.~\ref{fig:starFehistogram}, disc stars have an oxygen
content always larger with respect to bulge stars. Since these plots
show basically a mass-weighted metallicity, the oxygen abundance of
the total sample lies in between the values for disc and bulge stars.
Another important aspect is that the predicted values of the stellar
oxygen abundance are compatible with the observations although, if one
focuses on the total sample and on disc stars, it is at the upper end
of what it is currently observed, especially at large galactocentric
radii. This is presumably linked to the high $\rm{[O/Fe]}$ values at
high stellar metallicities that have been discussed in
Fig.~\ref{fig:oxygenvsironstars} and again suggests that in our
simulations the amount of oxygen may be mildly overproduced. 

We noted earlier that the oxygen profiles are declining with
radius; thus a metallicity gradient is present. For what concerns its
strength, it is apparent from the comparison with the two best-fitting
relations that in general the gradients predicted by our simulations
are too shallow, even considering the optical fit to the data which
has the least pronounced slope. Only in two cases (Aq-B-5 and Aq-C-5) the
slope of the optical best-fitting relation is reproduced, but there is
a considerable offset toward higher abundances, as previously
discussed.  In some of the haloes (Aq-A-5, Aq-B-5 and Aq-C-5), bulge stars
do a considerably better job in capturing the predicted metallicity
trend, both in slope and normalisation. 

We must caution the reader that \hii\ regions surround newly
formed stars in the Galactic disc and as such are associated to a very
young stellar population and the most metal rich gas. In our
comparison we did not use any age cut for the stars, thereby including
also old and metal poor stars. This in principle could affect our
conclusions. However, we have explicitly checked that even restricting
the comparison data to the youngest stars (age $<$ 300 Myr) does not
influence our results significantly. In some cases (Aq-A-5,
Aq-B-5 and Aq-C-5) this may lead to a steepening of the metallicity profile,
but a too flat stellar metallicity gradient is still present in the
majority of the simulated galaxies.

\section{Discussion} \label{sec:discussion}

In the previous sections we have presented the main properties of the
diffuse gas and stellar metallicity distributions that our simulations
predict for galaxies similar to our own Milky Way. Here we extend the
discussion on how these results compare to observations, examining in
more detail which observational constraints are matched and where
tensions between our simulations and the data lie.

We start with the properties of the diffuse gas surrounding the main
galaxy. The picture that emerges from the simulations is that the
structure of this gas component is strongly affected by galactic-scale
winds, which in our simulations are necessary to curtail star
formation. This is particularly evident if one compares the bipolar
morphology (caused by the fact that the galactic wind is
preferentially ejected perpendicularly to the star-forming disc in our
current implementation) in the column density map of metals to the
global morphology of the gas (see Fig.~\ref{fig:maps}). The same
comparison also shows that the diffuse gas is significantly
metal enriched out to and beyond the virial radius, and
Fig.~\ref{fig:metalposhist} backs up this conclusion. The average
metal content in the diffuse gas is a slowly declining function of
radius and, although a significant scatter is present, the average gas
metallicity is close to solar.  This value is larger than that
commonly inferred from observational constraints in the Milky Way
other nearby galaxies, which seem to prefer a value closer to $\sim
0.1\,Z_{\odot}$ \citep{Hodges-Kluck2013, Miller2013}.  However, due to
its very low density, the metallicity of the diffuse gas phase and
especially that of the hot emitting X-ray gas is poorly constrained.
This metallicity is hence often simply either assumed or allowed to
vary between $0.1-1\,Z_{\odot}$ \citep{Anderson2013}.

There is also another crucial factor that can substantially impact the
degree to which the halo gas is enriched by metals. In our wind model,
the metal loading of the wind can in principle be chosen
\textit{independently} from its mass loading. This was introduced to
accommodate the fact that the mass loading is expected to arise in part
through entrainment of gas with much lower metallicity than that of
the star-forming gas itself. Indeed, it proved necessary in our wind
model to allow galaxies to retain enough metals such that the
observational relation between their stellar masses and metal content
\citep[see Fig.~13 in][]{Vogelsberger2013} could be matched, while
still efficiently expelling baryons to control star formation. The
standard choice for our wind model is to load the outflow only with
$40\%$ of the metal mass contained in the parent gas cell, while
the remaining part is redistributed to the neighbouring ISM particles.
The results presented here suggest that even with this reduced metal
loading our galactic winds are able to significantly enrich
the CGM with metals and to eject an appreciable fraction of the
total metal mass beyond the virial radius of the enclosing
halo (see also Table~\ref{tab:metals}). Within 150 kpc our simulated haloes
contain 50-80\% of their total metal budget. Half of the sample is close
to the upper boundary of this range, which is a factor of $\sim 2$ higher
than what found in COS-Halos data \citep{Peeples2014}.

Although the distribution of metals in the halo is in tension with
what is observationally known, other properties of the diffuse gas
such as density and mass distributions match Milky Way and other
  late-type galaxies constraints quite well. The mass contained in
the diffuse gas component is a substantial fraction of the total
baryonic content of the halo (up to $\simeq 50\%$ of the universal
baryon fraction), but in general the haloes are not baryonically
closed (see last column of Table~\ref{tab:metals}). The total
densities are consistent with those determined at low redshift for
  $L_{*}$ galaxies \citep{Werk2014}. They are also compatible with the
  densities needed to strip gas via ram pressure from dwarf
spheroidal galaxies in the Milky Way \citep{Gatto2013}. However,
  we must note that the simulations in \citet{Gatto2013} were
  primarily aimed at investigating the hot phase of the diffuse CGM
  (the so-called hot corona). We caution again that this phase is
  somewhat underrepresented in our simulations, except in haloes
  Aq-A-5 and Aq-C-5.

\begin{figure*}
\resizebox{17.8cm}{!}{\includegraphics{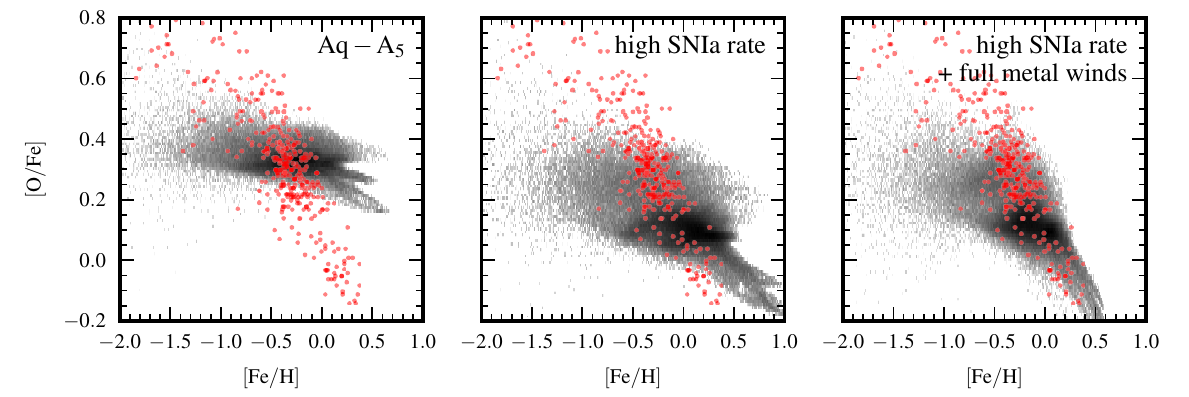}}
\caption{The same as Fig.~\ref{fig:oxygenvsironstars} but showing the
  $[{\rm O/Fe}]-[{\rm Fe/H}]$ plane for the Aq-A-5 halo with three
  different parametrization of type Ia SNae and wind metal loading:
  our fiducial set-up (left panel), a simulation with an increased
  normalization ($\sim 4$ times) in the time delay distribution of
  SNIa (middle panel) and a model that adds to the higher SNIa
  normalization fully metal-loaded winds. The increased SNIa rate in
  the middle and right panels results in a greater iron production
  that lowers the ${\rm [O/Fe]}$ values with respect to the fiducial
  case. Also, the anti-correlation between oxygen and iron content in
  the stellar component starts to emerge more clearly at high
  metallicities. In particular, the model that allows for fully
  metal-loaded outflows is now able to match the trend observed in the
  Milky Way. }
\label{fig:SNImodels}
\end{figure*}

The stellar metallicities in our simulations are able to capture the
general trends expected in the Galaxy. For instance, the stellar iron
content shows that disc stars are more metal rich and younger with
respect to their bulge counterpart and the iron distribution peaks
roughly at the solar value. However, when simulations are compared in
more detail to the observations, discrepancies start to emerge. We
note that in our model the processes for the injection of metals into
the ISM and for the ejection metals through wind particles are
related, but are not explicitly coupled as is frequently done in
feedback models that inject metals concurrently and into the same gas
as the energy that is intended to drive winds
\citep[e.g.][]{Agertz2011, Guedes2011, Aumer2013b, Simpson2013,
Stinson2013}. Therefore, the fidelity of our simulated stellar
metallicities with observations does not necessarily constrain the
previously described results for the properties of the CGM.

A significant discrepancy for stellar metallicities is that 
our simulations have only a very weak
dependence of the ${\rm [O/Fe]}$ abundance with respect to ${\rm
  [Fe/H]}$. This relation is an important indicator
of the relative contribution of core-collapse versus type Ia
supernovae to metal enrichment. As such, it represents an important
test of our stellar evolution module. In addition to the details of
the star formation history in each halo, there are two major factors
that can influence the ${\rm [O/Fe]}$-${\rm [Fe/H]}$ relation. The
first is due to the uncertainties in the metal yields that we
adopted in the model, which can induce an over- or under-production of
one or several of the heavy elements (for instance of oxygen as
mentioned in Sect.~\ref{sec:stars}). The second is the time delay
distribution of SNIa and its overall normalization relative to the
core collapse rate, which dictates when iron will be released into the
ISM and therefore can be incorporated into newly formed stars. All the
simulations discussed in this paper were performed with the fiducial
setup in \cite{Vogelsberger2013}, but in principle it is possible to
act on both factors to reconcile the simulation results with the
observational constraints.
Also, the efficiency with which metals are transported into the CGM by
galactic winds (which in our model is regulated by their metal
loading) can play a significant role. 

We explore these issues in Fig.~\ref{fig:SNImodels}, where we present
again the ${\rm[O/Fe]}$-${\rm [Fe/H]}$ relation for the Aq-A-5 halo
for three different models: our fiducial set-up (left panel), a
simulation which uses a higher ($\sim 4$ times) normalization of SNIa
rate (middle panel) and a model that in addition to the higher SNIa
rate allows for fully metal-loaded winds (right panel), that is
winds loaded with all the metal mass of their parent gas
cell\footnote{Note that also in the case of full metal loading the
ISM of the galaxy will contain metals, since the generation of wind
particles in our model is done on a cell-by-cell basis 
\citep[for details see][]{Vogelsberger2013}. The fact that
the ISM retains metals is also apparent from the right panel of
Fig.~\ref{fig:SNImodels}, which shows that the galactic stellar
component has stars with solar or slightly super-solar iron
content.}. The suggested increase in the normalization of the SNIa
rate contains two factors, one is from changing the fiducial value of
the model parameter from $1.3\times 10^{-3}$ \citep{Maoz2012} to
$2.5\times 10^{-3}\,[\rm SN\,M_{\odot}^{-1}]$, and another factor of
about 2 arises from re-defining the SNIa rate parameter as the total
number of SNIa per solar mass that explode in the stellar population
over a Hubble time ($\simeq 13.56\,\Gyr$ for this simulation set). In
\citet{Vogelsberger2013}, instead, the upper limit of the
normalization integral for the SNIa delay time distribution was taken
as infinite. Due to the shallowness of the adopted power law for the
SNIa time delay distribution (the power law index is $s = -1.12$),
there is then approximately a factor of 2 between these normalization
conventions.

From Fig.~\ref{fig:SNImodels} it can be seen that
the increased number of SNIa events in the last two models leads to a
higher iron production and therefore to smaller values of the ${\rm
  [O/Fe]}$ abundance. More interestingly, a clear anti-correlation
between the iron and oxygen content starts to appear at the
high-metallicity end. In particular, the model that allows for fully
metal-loaded winds provides an impressively good match of the data for
the Milky Way. We explicitly checked that the models with increased
SNIa rates yield consistent results for what concerns the galaxy
properties examined in \citetalias{Marinacci2013}. We found no
significant systematic differences except a somewhat higher SFR at
late times (an expected trend due to the enhanced cooling of the gas
caused by the slightly higher metal production of these simulations)
that makes the galaxies appear slightly bluer than in our default run.

The other important mismatch between our simulations and the
observations is the presence of only a shallow radial metallicity
gradient in the stellar component. A possible explanation for this
result is that our discs may suffer from an excessive radial migration
of stellar orbits that can considerably reduce any pre-existing radial
metallicity gradient, an issue that we will fully explore in
future work (Yurin et al., in prep.). An alternative explanation to
be considered is that the material from which stars are born is too
metal-rich, as a consequence of the strong stellar feedback
included in our calculations. It is fairly well established by
chemical evolution models of our Galaxy \citep[e.g.,][]{Chiappini2001}
that accretion of low-metallicity gas is needed to sustain its current
SFR and to match its chemical signatures. In our simulations discs
form in an inside-out fashion with gas accretion occurring
preferentially at disc edges \citepalias[see Figs.~9 and 10
in][]{Marinacci2013}. In such a scenario, the accretion of relatively
metal-poor material would naturally lead to a gradient in the star
metallicities if radial migration of stars is not particularly strong.
However, as we have discussed above, the material outside the disc has
roughly solar metallicity and this similarity in chemical composition
with the stellar component might cause a considerable flattening of
the expected metallicity gradient. This explanation about the
flatness of the stellar metallicity gradient is in line with what
is discussed by \citet{Gibson2013} in cosmological simulations of galaxy
formation of Milky Way-type objects. In their sample of simulated
galaxies these authors find that strong stellar feedback models lead
to flat metallicity gradients, an effect that they ascribe to the
effectiveness of strong feedback in enriching the CGM and in
redistributing this metal-enriched material -- that fuels late-time
star formation -- on large scales.

\section{Summary and conclusions} \label{sec:conclusions}

In this paper, we have investigated the properties of the CGM 
and \FM{the metal content} of the stars comprising the
central galaxy in eight cosmological `zoom-in' simulations of Milky
Way-sized haloes taken from the `Aquarius' initial condition set. The
simulations were performed with the moving-mesh code \arepo\ combined
with a comprehensive model for galaxy formation physics, which
includes, in addition to galactic winds, metal cooling, a
self-consistent treatment of the stellar evolution and the associated
mass and metal return to the gaseous phase, as well as AGN feedback.
The simulations lead to the formation of realistic disc-dominated
galaxies whose properties agree quite well with the scaling relations
and the observed features of late-type systems. The main findings of
our analysis can be summarized as follows:

\begin{enumerate}
\item galactic winds are the main channel for metal enrichment of the
  CGM as can be inferred from the bipolar morphology (induced by our
  wind model) of the distribution of the heavy elements in the halo;
\item the circum-galactic regions of all the simulated galaxies are
  substantially metal enriched by the winds \FM{that in some cases can transport
  up to $40\%$ of the total metals beyond the virial radius of the enclosing halo};
\item the average mass-weighted
  metallicity of the CGM is $\sim Z_{\odot}$ and shows a slowly declining trend with
  radius;
\item the density profile and the total mass of the diffuse
  circum-galactic gas are in agreement with Galactic and $L_{*}$
  galaxies observational
  constraints and the mass contained in this gas reservoir is a
  significant fraction of the baryon budget of the halo, although it
  is in general not sufficient to make the halo baryonically closed;
\item winds also alter the thermal state of the CGM and supply an
  energy input capable of sustaining its radiative losses;
\item disc and bulge stars show a rather different metallicity distribution
  with disc stars more metal rich and more narrowly distributed around their metallicity
  peak (at about $Z_{\odot}$) than bulge stars;
\item stars in the central galaxy exhibit a declining metallicity profile
  as a function of galactocentric radius, however this gradient is too weak 
  when compared to Galactic observations;
\item the relation between oxygen over iron abundance and metallicity
  in the stellar component shows a trend which is too shallow with
  respect to Galactic observations;  the oxygen vs
  iron abundance can be reconciled with observations by slightly
  changing the parameters describing the time delay distribution of
  SNIa events and the efficiency with which winds transport metals
  from the central galaxy to the CGM.
\end{enumerate}

It is worth pointing out that the free parameters of our galaxy
formation physics model, which are primarily those determining the
strength of the galactic winds, were set to match the final stellar
mass of the galaxy and not the properties of the diffuse
circum-galactic gas and its metal enrichment properties or those
of stars comprising the central galaxy. Therefore, the results listed
above are genuine predictions of our galaxy formation model for these
properties in late-type galaxies, which can thus be used to check the
consistency of the model with constraints that are independent from
those considered to set the parameters. The outcome of the analysis
carried out in this work suggests that our model can satisfactorily
predict the general properties of the circum-galactic medium
associated with late-type systems and the metal content of stars
contained within the central galaxy. However, at a more detailed
level of comparison with the observations discrepancies with the data
start to emerge. These discrepancies are signaling deficiencies in our
parameterization of the supernova type Ia rate, and possibly also in
our phenomenological galactic wind implementation. Resolving these
tensions in refined theoretical simulation models will be very helpful
in further clarifying the connection between galaxies and their
circum-galactic environment as well as for understanding the
implications of this connection for galactic evolution.

\section*{Acknowledgements}

We thank an anonymous referee for her/his thoughtful comments and
suggestions. We also thank Andrea Gatto for making his routines to
generate coronal profiles available to us and Lars Hernquist, Mark
Vogelsberger and Till Sawala for useful discussions. FM and VS
acknowledge support by the DFG Research Centre SFB-881 `The Milky Way
System' through project A1. This work has also been supported by the
European Research Council under ERC-StG grant EXAGAL-308037 and by the
Klaus Tschira Foundation. Part of the simulations of this paper used
the superMUC system at the Leibniz Computing Centre, Garching, under
the project PR85JE of the Gauss Centre for Supercomputing, Germany.

\bibliographystyle{mn2e}
\bibliography{paper}

\label{lastpage}

\end{document}